\documentclass[3p,times,twocolumn]{elsarticle}

\usepackage{ecrc}

\volume{00}

\firstpage{1}

\journalname{Nuclear Physics B Proceedings Supplement}

\runauth{Sergey Alekhin, Johannes Bl\"umlein, Sven Moch}

\jid{nuphbp}

\jnltitlelogo{Nuclear Physics B Proceedings Supplement}

\usepackage{amssymb}
\usepackage[figuresright]{rotating}

\def\MSbar{{$\overline{\mbox{MS}}\,$}}

\begin{document}

\begin{frontmatter}



\dochead{}

\title{{\small DESY 12-030, DO-TH 12/08, SFB/CPP-12-14, LPN 12-03 
\hfill {\tt arXiv:1202.xxxx[hep-ph]}}
\\PDF fit in the fixed-flavor-number scheme}


\author{S.~Alekhin$^{\, a,b,}$,
    J.~Bl\"umlein$^{\, a,}$,
    and S.~Moch$^{\, a,}$ }

\address  {\it
    $^a$Deutsches Elektronensynchrotron DESY \\
    Platanenallee 6, D--15735 Zeuthen, Germany \\
    \vspace{0.2cm}
    $^b$Institute for High Energy Physics \\
    142281 Protvino, Moscow region, Russia\\
  }

\begin{abstract}
We discuss the heavy-quark contribution to deep inelastic scattering in the scheme with $n_f=3,4,5$ fixed flavors.
Based on the recent ABM11 PDF analysis of world data for deep-inelastic scattering 
and fixed-target data for the Drell-Yan process with the running-mass
definition for heavy quarks 
we show that fixed flavor number scheme is sufficient for describing the
deep-inelastic-scattering data in the entire kinematic range.
We compare with other PDF sets and comment on the implications for measuring
the strong coupling constant $\alpha_s(M_Z)$.
\end{abstract}


\end{frontmatter}

% \linenumbers


Account of the heavy-quark contribution to deep-inelastic scattering (DIS)
is an important issue for the PDF fits. 
The charm- and bottom-quark contributions, which are mostly relevant 
for the existing data analysis, 
are substantial at small values of the Bjorken variable $x$ forming
an important contributions to 
the HERA data on the inclusive structure functions (SFs). 
Furthermore, the semi-inclusive SFs, which correspond to the subprocess 
with the heavy quarks in the final state,  
can provide an additional constraint on the small-$x$ gluon 
distribution~\cite{Witten:1975bh}. However, 
the calculation of the heavy-quark production cross section
is hampered within perturbative QCD 
because the higher-order corrections are extremely involved for the case of two 
scales, the lepton momentum transfer $Q^2$ and the heavy-quark mass.
At present the QCD corrections to the massive Wilson coefficients
are known up to the NLO 
only~\cite{Laenen:1992zk, Gottschalk:1980rv, Gluck:1996ve}.
The partial NNLO corrections stemming from the soft-gluon
threshold re-summation 
have been also calculated for the neutral-current (NC)~\cite{Laenen:1998kp} and 
charged-current (CC) heavy-quark production~\cite{Corcella:2003ib}.
For the case of electro-production they are numerically important 
at small $x$ and $Q^2$ reaching O(10\%) 
in the kinematic region of  HERA. 
The NNLO threshold re-summation calculations
of Ref.~\cite{Laenen:1998kp} were recently 
updated~\cite{Alekhin:2008hc, Presti:2010pd} and now they include 
all threshold-enhanced logs and the Coulomb term.
The heavy-quark mass appearing in the massive Wilson coefficients 
of Refs.~\cite{Witten:1975bh, Laenen:1992zk, Gluck:1996ve, Presti:2010pd}
corresponds to the pole mass which emerges in the QCD Lagrangian. 
The pole-mass definition provides a straightforward way for the perturbative
calculations, however, it is not ideal for phenomenology  
since the pole mass is quite sensitive to the QCD radiative corrections. 
This shortcoming is eliminated when the heavy-quark mass is defined 
in the \MSbar-scheme, similarly to the strong coupling 
constant $\alpha_s$. The heavy quark 
\MSbar-masses are conventionally parametrized at the scales of 
the heavy-quark mass itself.
These scales are close to the typical DIS hard-scattering scale therefore 
their perturbative stability is greatly enhanced if compared to the 
pole-mass definition. The massive
Wilson coefficients re-calculated in terms of  
the running mass also demonstrate improved
perturbative stability and the reduced renormalization/factorization scale 
sensitivity at the HERA kinematics~\cite{Alekhin:2010sv}. 
In this range only the charm- and 
bottom-quark production are relevant. The $c$- and
\begin{figure*}[th!]
\centerline{
  \includegraphics[scale=0.4]{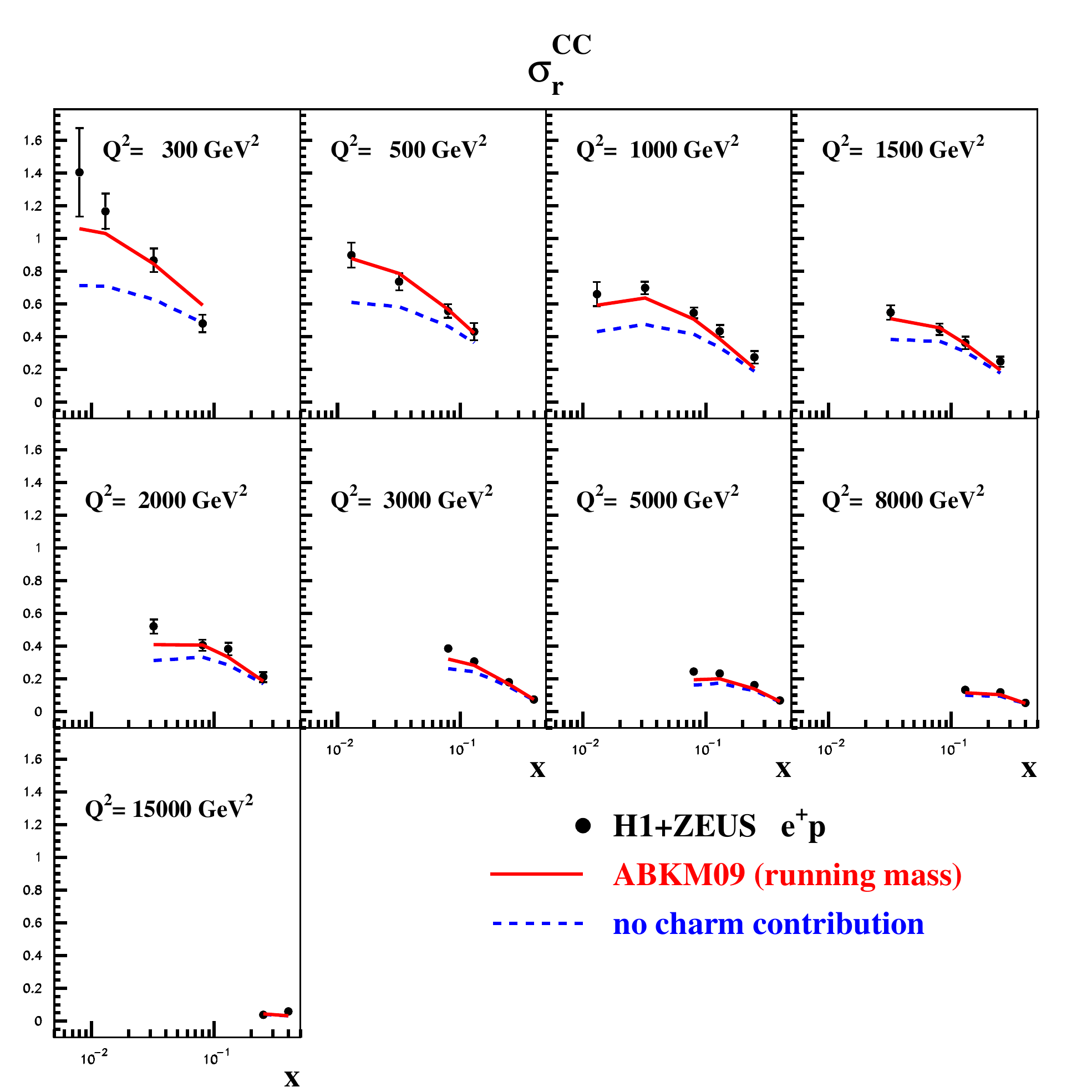}
  \includegraphics[scale=0.4]{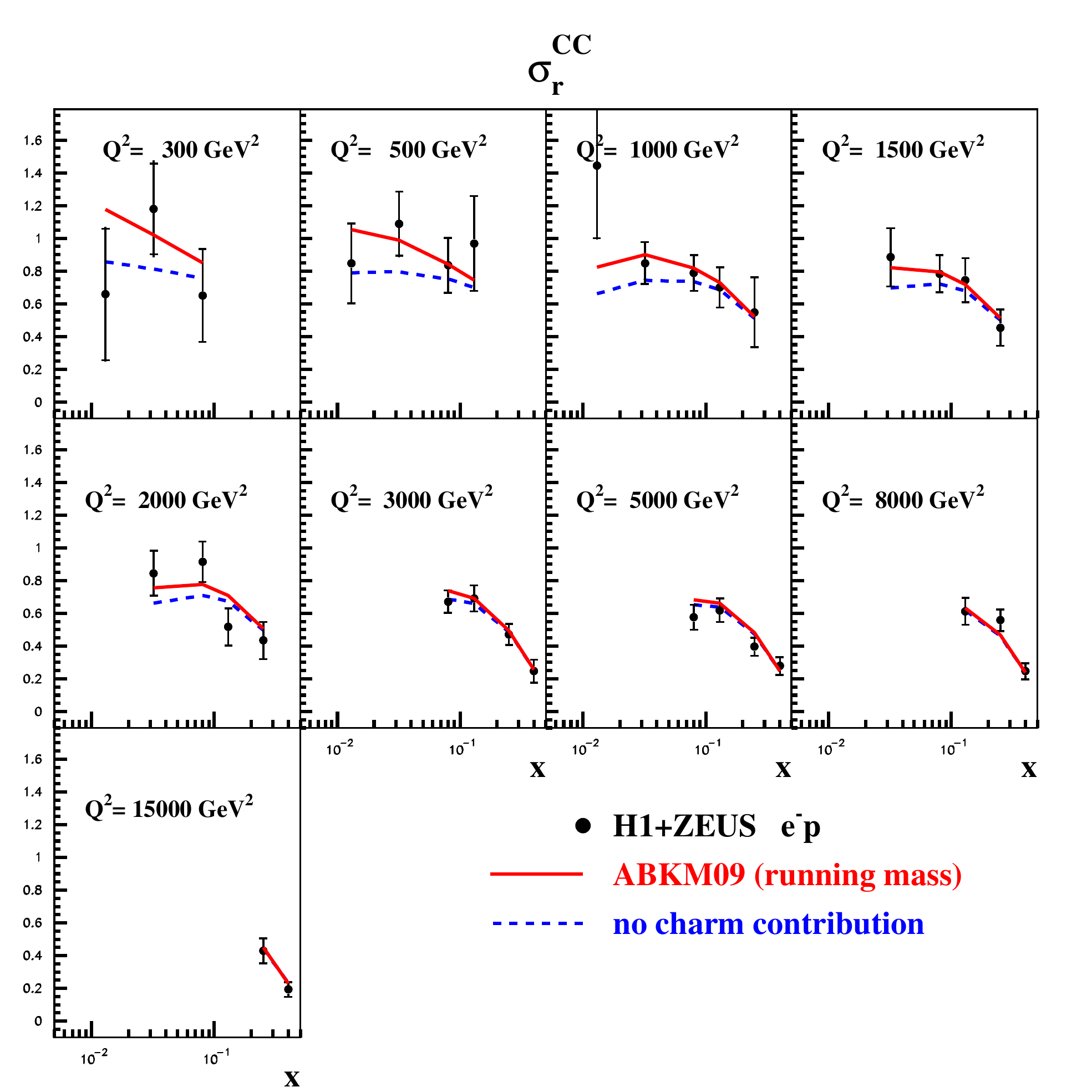}}
   \caption{\small
    \label{fig:cc}
     The CC inclusive cross section data
     obtained at HERA~\cite{:2009wt} with the positron beam (left)
     and the electron beam (right)
     compared with the running-mass 3-flavor scheme predictions based on the 
     NNLO PDFs of Ref.~\cite{Alekhin:2010sv} (solid lines). 
     The same predictions without 
     the $c$-quark contribution taken into account
      are given for comparison (dashes). 
}
\end{figure*}
$b$-quark
\MSbar-masses averaged over different determinations are 
\begin{eqnarray}
  \label{eq:mcmbinp}
  m_c(m_c) \,=\, 1.27 \pm 0.08\,\, {\rm GeV}
\end{eqnarray}
and
\begin{eqnarray}
  \label{eq:mbmbinp}
  m_b(m_b) \,=\, 4.19 \pm 0.13\,\, {\rm GeV}
  \, ,
\end{eqnarray}
respectively~\cite{Nakamura:2010zzi}. 
The uncertainties in these values are essentially smaller 
than the spread in the heavy-quark masses values employed in 
the various global PDF fits, cf. Tab.~1 in~\cite{Alekhin:2010sv}.
Therefore the running-mass definition provides a footing for 
the consolidation of those PDFs. 

Taking advantage of the running-mass definition and 
new massive NNLO corrections we update our ABKM09 
PDFs, which were obtained within the fixed-flavor-number (FFN) scheme 
at NLO and NNLO~\cite{Alekhin:2009ni}, 
and produce in this way a new PDF set, ABM11.  
Apart from this theoretical improvement we add to the fit
recently published HERA data. 
Firstly, the separate H1~\cite{Adloff:2000qk} 
and ZEUS~\cite{Chekanov:2001qu} data sets on the NC inclusive SFs 
are replaced by much more accurate combined HERA data~\cite{:2009wt}.
The unprecedentedly small uncertainty of 1-2\%, including the 
normalization error of 0.5\%, achieved 
for the combined HERA inclusive data allows to reach better constraints on the 
small-$x$ PDFs as compared to the ABKM09 analysis.
Only the NC combined HERA data with $Q^2<1000~{\rm GeV}^2$ are used in order
to exclude contributions due to the $Z$-boson exchange without 
loosing the statistical significance of the NC HERA data for our fit. 
Besides, we include into the fit the CC HERA data obtained 
for the electron and positron beams~\cite{:2009wt}. The CC 
HERA sample ranges up to $Q^2=15000~{\rm GeV}^2$. To demonstrate
that the treatment of the charm production within the
FFN scheme is applicable at that large transverse 
momentum we compared the 
predictions based on the variant of the NNLO ABKM09 fit performed with the 
running-mass scheme~\cite{Alekhin:2010sv}.
The value of $m_c(m_c)=1.18\pm0.06~{\rm GeV}$ is obtained in this fit 
with the constraint of Eq.~(\ref{eq:mcmbinp}) imposed. 
Taking this value of $m_c(m_c)$ we find very good agreement 
with the CC HERA data in the whole range of $Q^2$, 
cf. Fig.~\ref{fig:cc}. At small $x$ the charm contribution to the inclusive 
CC cross sections is quite significant, similarly to the 
case of the NC DIS. Since the CC charm production 
is initiated by the strange quarks mainly, this sample provides an
additional constraint on the small-$x$ strange sea distribution, which 
 is at the moment defined by the neutrino-nucleon DIS data 
only~\cite{Alekhin:2008mb}.
In the same way the charged-lepton initiated CC data can be used to determine 
the value of the $c$-quark mass, in particular employing the potential 
of the planned high-luminosity EIC facility~\cite{Boer:2011fh}. 
Finally, we add to the fit the inclusive H1 data obtained in a special HERA  
run with high inelasticity $y$ achieved~\cite{Collaboration:2010ry}.
At the kinematics probed in 
\begin{figure}[h!]
\centerline{
  \includegraphics[width=8cm, height=12cm]{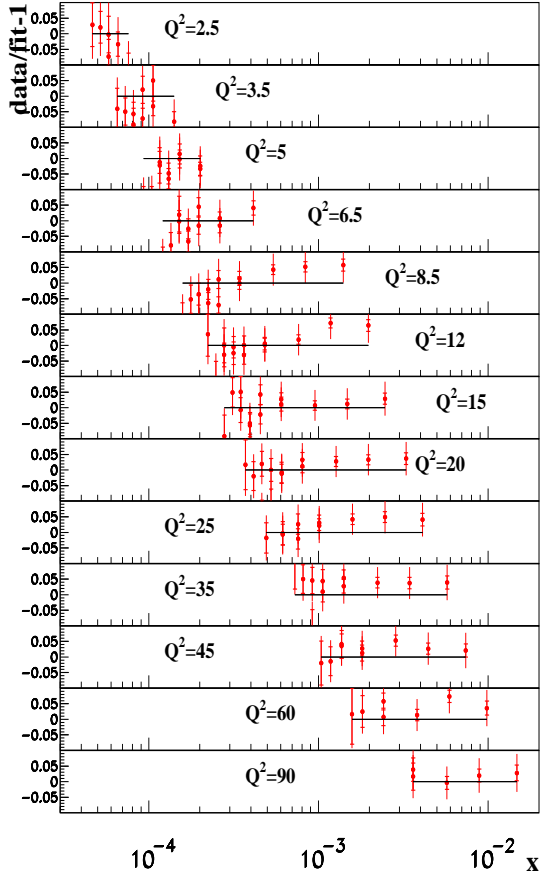}}
  \caption{\small
    \label{fig:heralow}
      The pulls versus $x$ for 
      the H1 NC inclusive DIS c.s.
      data of Ref.~\cite{Collaboration:2010ry}
      binned on the momentum transfer $Q^2$  in units of ${\rm GeV}^2$
      w.r.t. the ABM11 NNLO fit.
      The data points with different inelasticity $y$ still
      may overlap in the plot.
      The inner bars show statistical errors in data and the outer bars the
statistical and systematical errors combined in 
quadrature~\cite{Alekhin:2012ig}.
}
\end{figure}
\begin{figure}[th!]
\centerline{
  \includegraphics[scale=0.4]{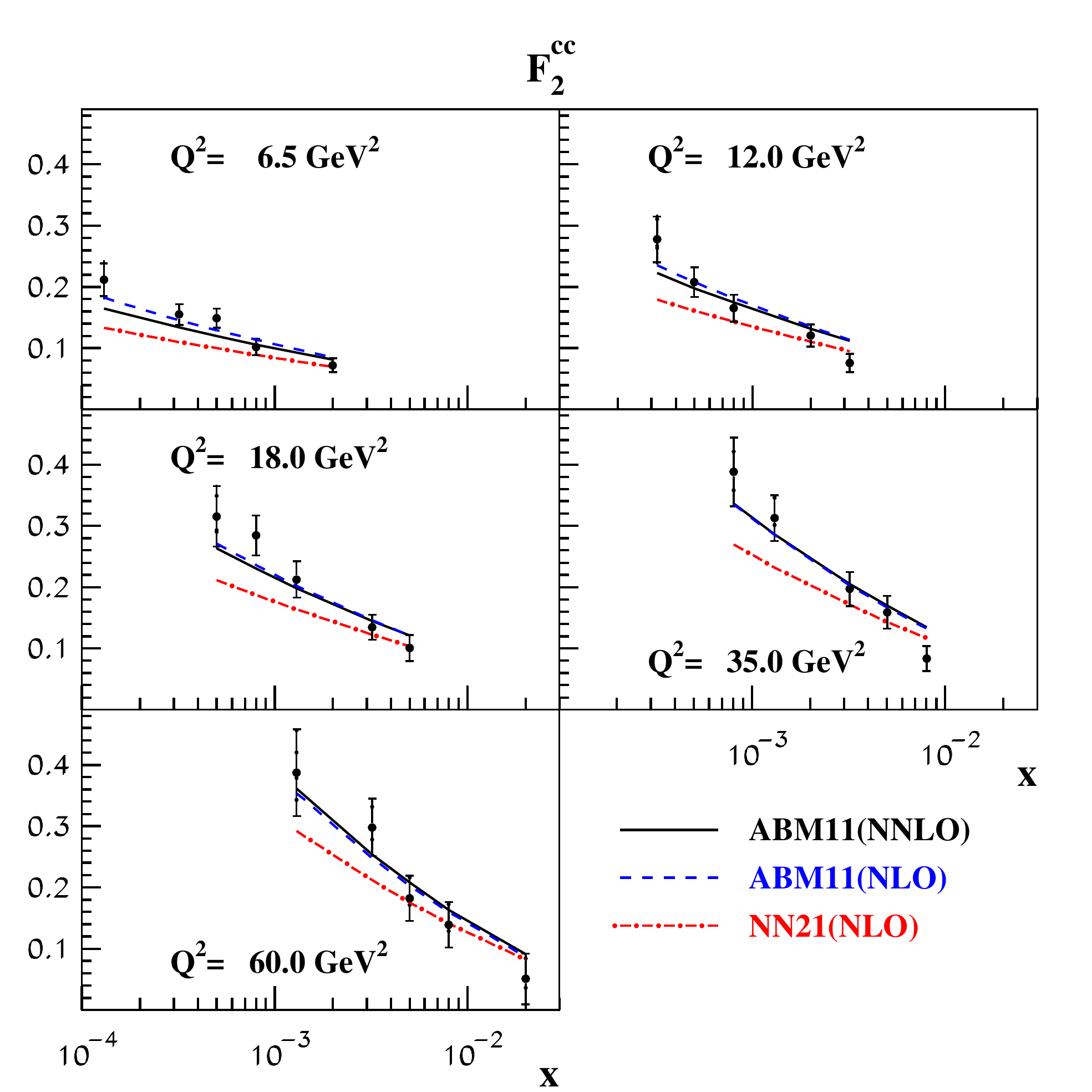}}
  \caption{\small
    \label{fig:h1cc}
      The predictions for the semi-inclusive structure function
      $F_2^{cc}$ at different values of the momentum transfer $Q^2$ versus $x$
      for the 
      ABM11 PDFs in the NNLO (solid curves), the ABM11 PDFs in the NLO 
      (dashes), and the NN21 PDFs~\cite{Ball:2011mu} in the NLO 
      (dashed dots), all  
      taken in the 3-flavor running-mass scheme with the
      the value of $m_c=1.27~{\rm GeV}$~\cite{Nakamura:2010zzi}. 
      The NNLO predictions based on the 
      MSTW~\cite{Martin:2009iq} and JR09~\cite{JimenezDelgado:2008hf} PDFs (not 
      displayed in the plot) strongly overlap with the ABM11 ones. 
      The H1 data of Ref.~\cite{Aaron:2011gp} given in the plot
      are extracted with HVQDIS code of Ref.~\cite{Harris:1995tu} .
}
\end{figure}
this run the inclusive SFs are 
sensitive to the structure function $F_L$ at small 
$x$ and $Q^2$ and thereby shed light on the details of the small-$x$ 
QCD dynamics, in particular the shape of the gluon distribution at small $x$. 

Similar to the ABKM09 analysis, in ABM11 the HERA data are supplemented by the 
ones on dimuon production in the 
(anti)neutrino-nucleon DIS and the Drell-Yan process in order to 
separate the quark PDFs by flavors. Also we employ the
fixed-target inclusive DIS data obtained in the  
NMC~\cite{Arneodo:1996qe}, 
the BCDMS~\cite{Benvenuti:1989rh,Benvenuti:1989fm}, 
and the SLAC experiments~\cite{Whitlow:1990gk,Bodek:1979rx,Atwood:1976ys,Mestayer:1982ba,Gomez:1993ri,Dasu:1993vk}. The fixed-target data  
allow to constrain the PDFs at large $x$. However, 
modeling of the higher-twist terms
is required since they contribute to the DIS SFs at small
$Q^2$~\cite{Virchaux:1991jc}. 
The corrections for nuclear effects~\cite{Kulagin:2004ie}
are also taken into account in the analysis of fixed target data when 
relevant, cf.~\cite{Alekhin:2012ig} for details.
The overall quality of the fit is quite good.
For 3036 data points used in the fit the value of $\chi^2$ is 
3391 and 3378 for the NLO and NNLO variants, respectively. 
For the combined HERA
data~\cite{:2009wt} including the NC and CC samples
the value of $\chi^2/NDP=537/486$ is obtained at NNLO  
with full account of the systematics error correlations.
The pulls of NC HERA data w.r.t. the fit
do not demonstrate a statistically significant trend versus $Q^2$.
This gives additional justification for 
application of the FFN scheme to the analysis 
of the DIS data at realistic kinematics, cf. also~\cite{Gluck:1993dpa}.
At small $x$ and $Q^2$ the perturbative QCD corrections rise and 
they are particularly big for the NNLO corrections to
the massless Wilson coefficients for $F_L$~\cite{Vermaseren:2005qc}.
With account of these corrections 
the high-$y$ H1 data~\cite{Collaboration:2010ry}, which 
are quite sensitive to the contribution from $F_L$
are described pretty well, with the value of $\chi^2=137$ for 130 data points.
Also the H1 data do not suggest a violation of the conventional QCD 
evolution down to $x\sim 10^{-5}$, cf. Fig.~\ref{fig:heralow}. 
In this way we do
\begin{figure}[th!]
\centerline{
  \includegraphics[scale=0.4]{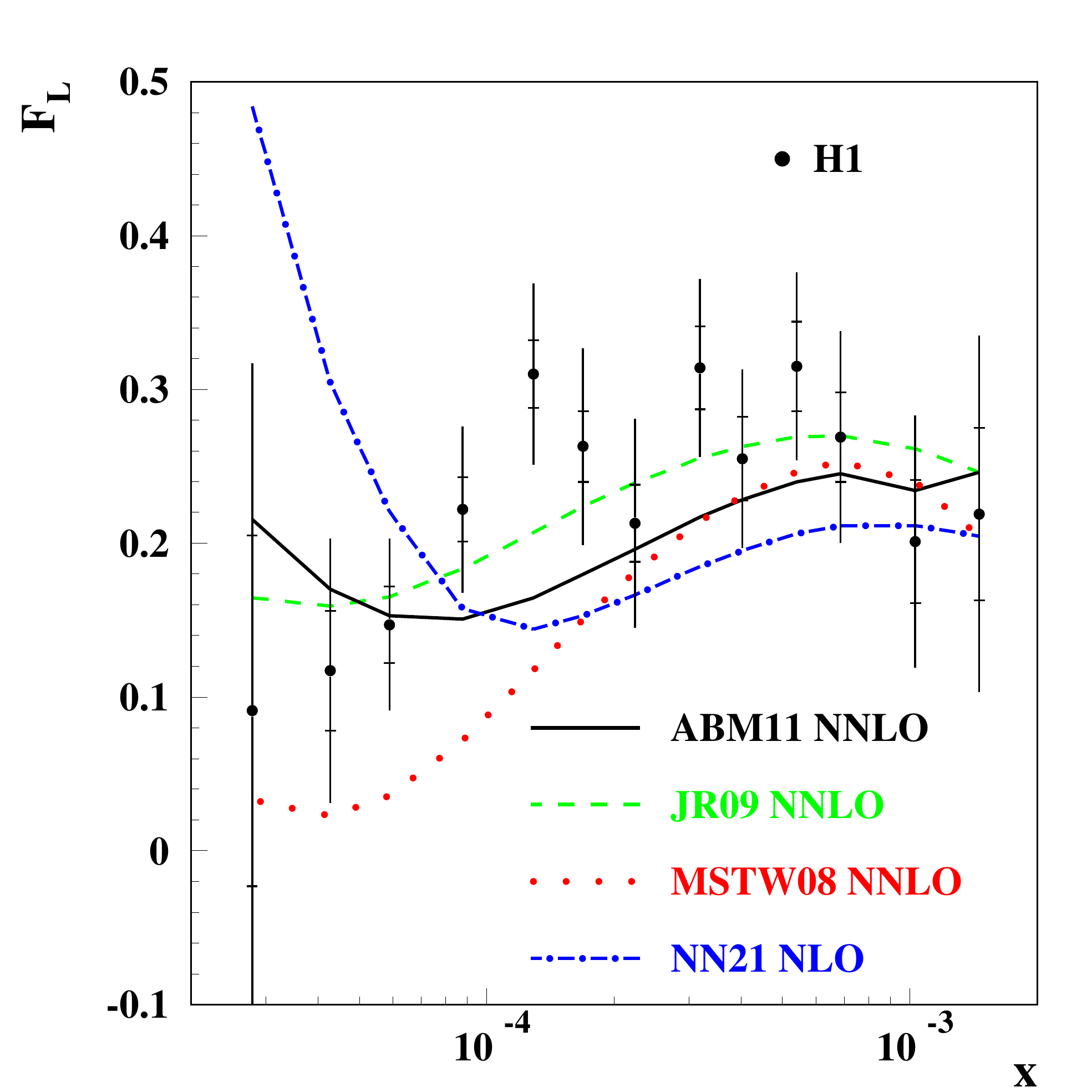}}
  \caption{\small
    \label{fig:flh1}
     The data on $F_L$ versus $x$ obtained by the H1
      collaboration~\cite{Collaboration:2010ry} 
     confronted with the 3-flavor scheme NNLO predictions based on the 
      different PDFs (solid line: this analysis, dashes: 
      JR09~\cite{JimenezDelgado:2008hf}, dots: MSTW08~\cite{Martin:2009iq}). 
      The NLO predictions based on the 3-flavor NN21 
       PDFs~\cite{Ball:2011mu} are given for comparison (dashed dots). 
       The value of $Q^2$ for the data points and the curves in the plot 
       rises with $x$ in the range of 
$1.5\div45~{\rm GeV}^2$~\cite{Alekhin:2012ig}.
}
\end{figure}
\begin{figure}[th!]
\centerline{
  \includegraphics[scale=0.4]{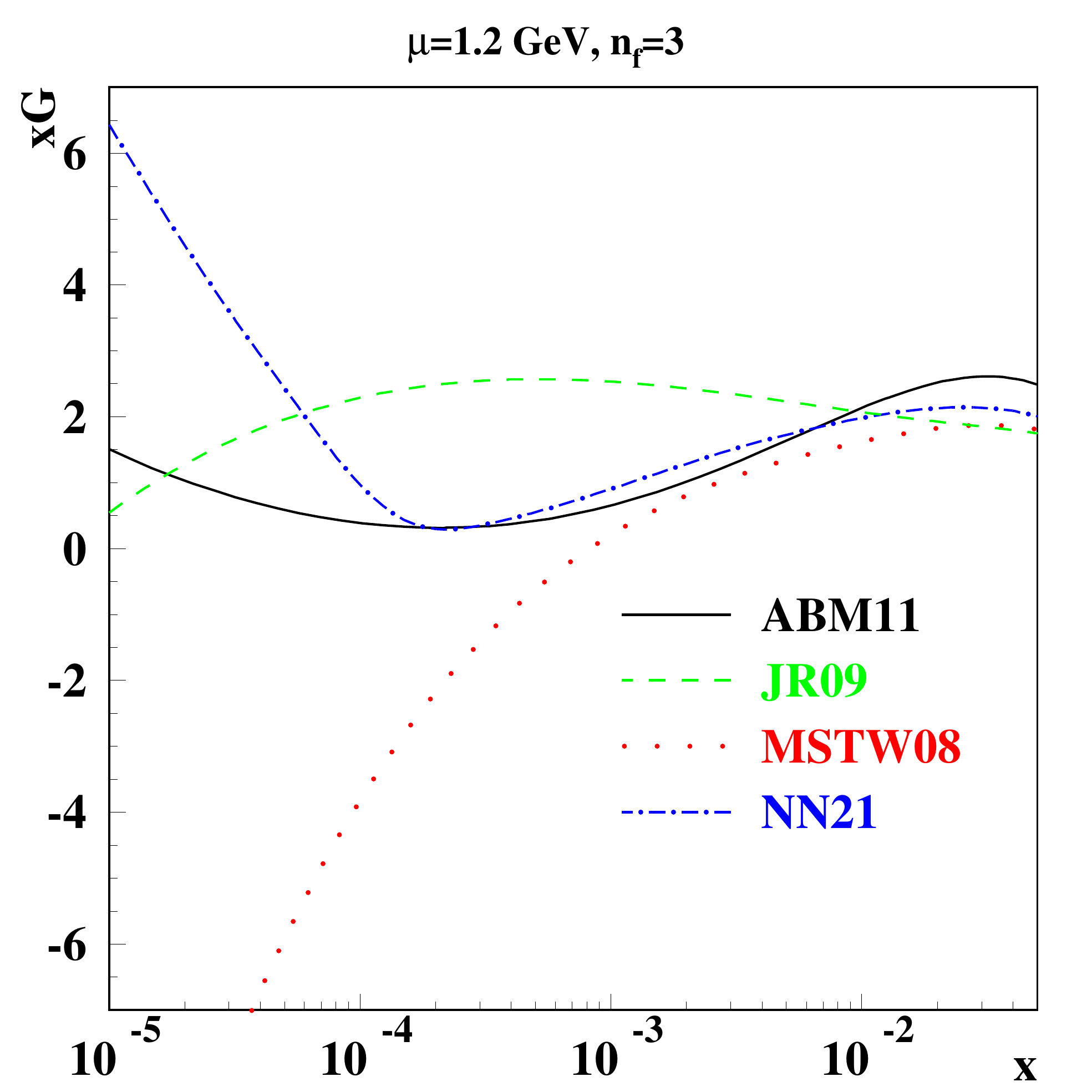}}
  \caption{\small
    \label{fig:lowg}
      The 3-flavor gluon distributions 
      obtained in the various 
      NNLO fits at the scale of $\mu=1.2~{\rm GeV}$
      (solid line: this analysis, dashes: 
      JR09~\cite{JimenezDelgado:2008hf}, dots: MSTW08~\cite{Martin:2009iq},
      dashed dots: NN21~\cite{Ball:2011uy}).    
}
\end{figure}
not confirm a hint on 
the evolution kernel re-summation effects observed in the 
NNPDF fit~\cite{Caola:2009iy}. It is worth noting
in this connection that the interpretation of the low-$x$ DIS inclusive data 
is sensitive to the treatment of the heavy-quark contribution. 
While we employ in the analysis of DIS data 
the FFN scheme with 3 flavors, in many other PDF 
fits~\cite{Martin:2009iq,Ball:2011uy,Lai:2010vv} 
the variable-flavor-number (VFN) scheme is used.
The VFN approach to the description of the DIS is based on the 
asymptotic expressions for massive Wilson coefficients
taken in the 3-flavor FFN scheme~\cite{Buza:1996wv,Bierenbaum:2009zt}, 
which are valid at $Q^2\gg m_h^2$, where $m_h$ is the heavy-quark mass. 
In this limit the power corrections vanish and the structure  
functions come down to convolutions of the massless Wilson coefficients 
with the PDFs, which now are defined in the 4-flavor scheme including 
also the heavy-quark PDFs. 
The matching conditions between 3- and 4-flavor schemes are known up to 
the second order in  
$\alpha_s$~\cite{Buza:1996wv,Bierenbaum:2009zt} 
and the first third-order corrections are also 
available~\cite{Bierenbaum:2009mv,Ablinger:2010ty}. However, in order to 
employ the VFN scheme in the analysis of realistic data, which range
down to $Q^2 \sim m_h^2$, one has to take into account 
the power corrections as well.  
This task is conceptually difficult within the VFN formalism therefore
the power corrections are commonly modeled using the combination of 
the massive FFN and massless VFN Wilson coefficients~\cite{Aivazis:1993pi}. 
The models which 
are obtained in this way pretend to describe the full-range kinematics
of existing DIS data; 
therefore they should reproduce the 3-flavor scheme results at small $Q^2$. 
A particular shape of the massive VFN coefficient functions at $Q^2 \ngg m_h^2$
is subject to a particular choice and there are
numerous VFN scheme prescriptions, which differ basically by 
the degree of smoothness provided for matching with the FFN scheme. 
In cases the smoothness is achieved by introducing
empirical parameter(s), which control dumping of the massless VFN
term at small $Q^2$~\cite{Thorne:2012qh}. It is worth noting that 
the uncertainty in a 
particular choice of these parameters is propagated into the uncertainties 
in PDFs obtained with such prescriptions. Moreover, 
additional parameters, which appear in the coefficient functions 
of Ref.~\cite{Thorne:2012qh} do not enter into the 
QCD anomalous dimensions.
Therefore the QCD factorization may be broken in this way.  
In contrary, the NNLO S-ACOT-$\chi$ prescription~\cite{Guzzi:2011ew} 
is explicitly based on the factorization theorem and does not contain 
damping factors.
However the matching smoothness is not guaranteed in this case
and the S-ACOT-$\chi$ based calculations 
overshoot the FFN scheme results at $Q^2 \sim m_h^2$, cf. Fig.~4 in 
~\cite{Guzzi:2011ew}.
On the 
other hand, the BMSN prescription~\cite{Buza:1996wv} 
ensures smooth matching 
with the FFN scheme at $Q^2 = m_h^2$ inherently, without additional 
parameters. Furthermore
the results of the PDF-analysis are not sensitive to the 
choice of either the VFN scheme as the BMSN prescription or the FFN 
scheme~\cite{Alekhin:2009ni}. 

\begin{figure*}[t]
  \includegraphics[scale=0.4]{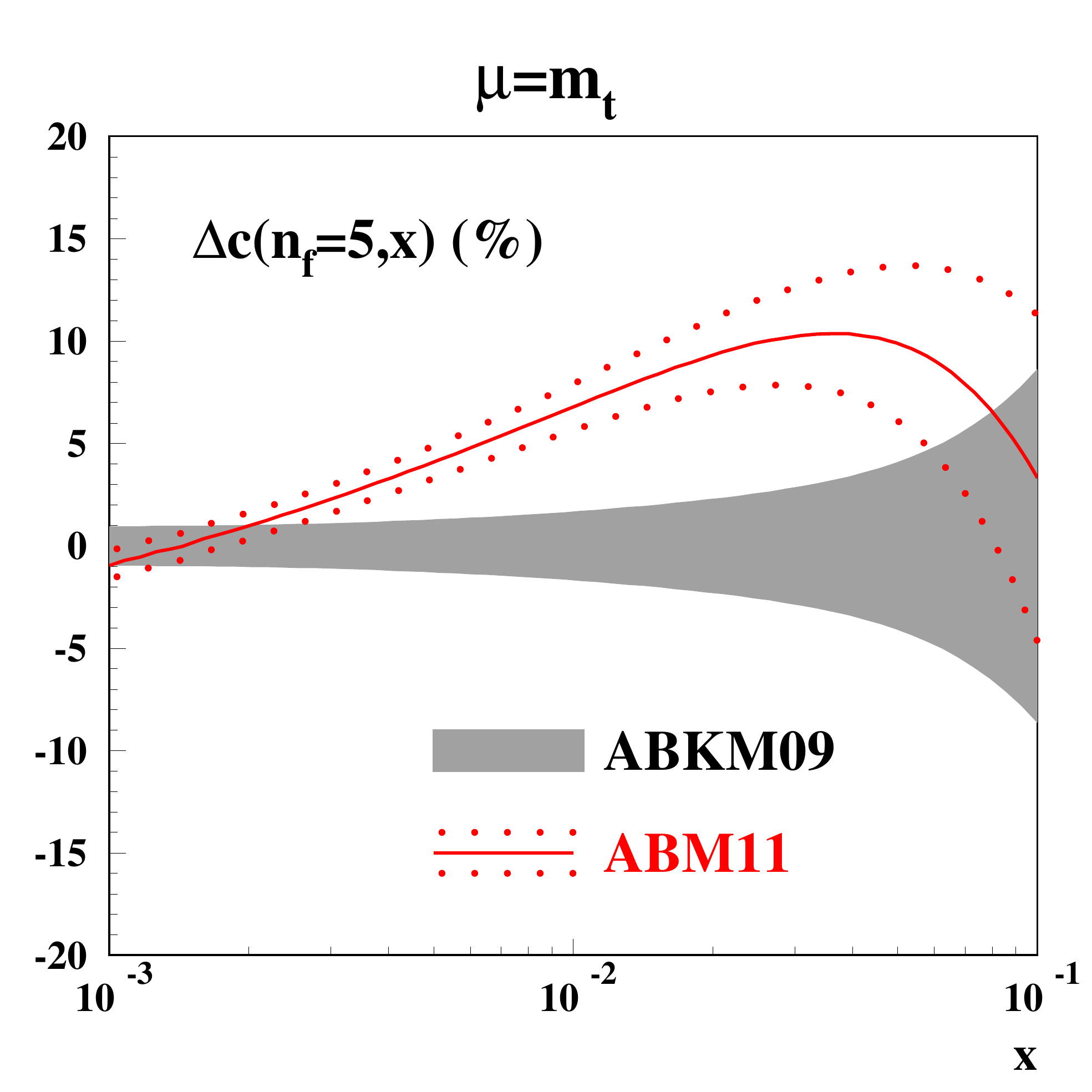}
  \includegraphics[scale=0.4]{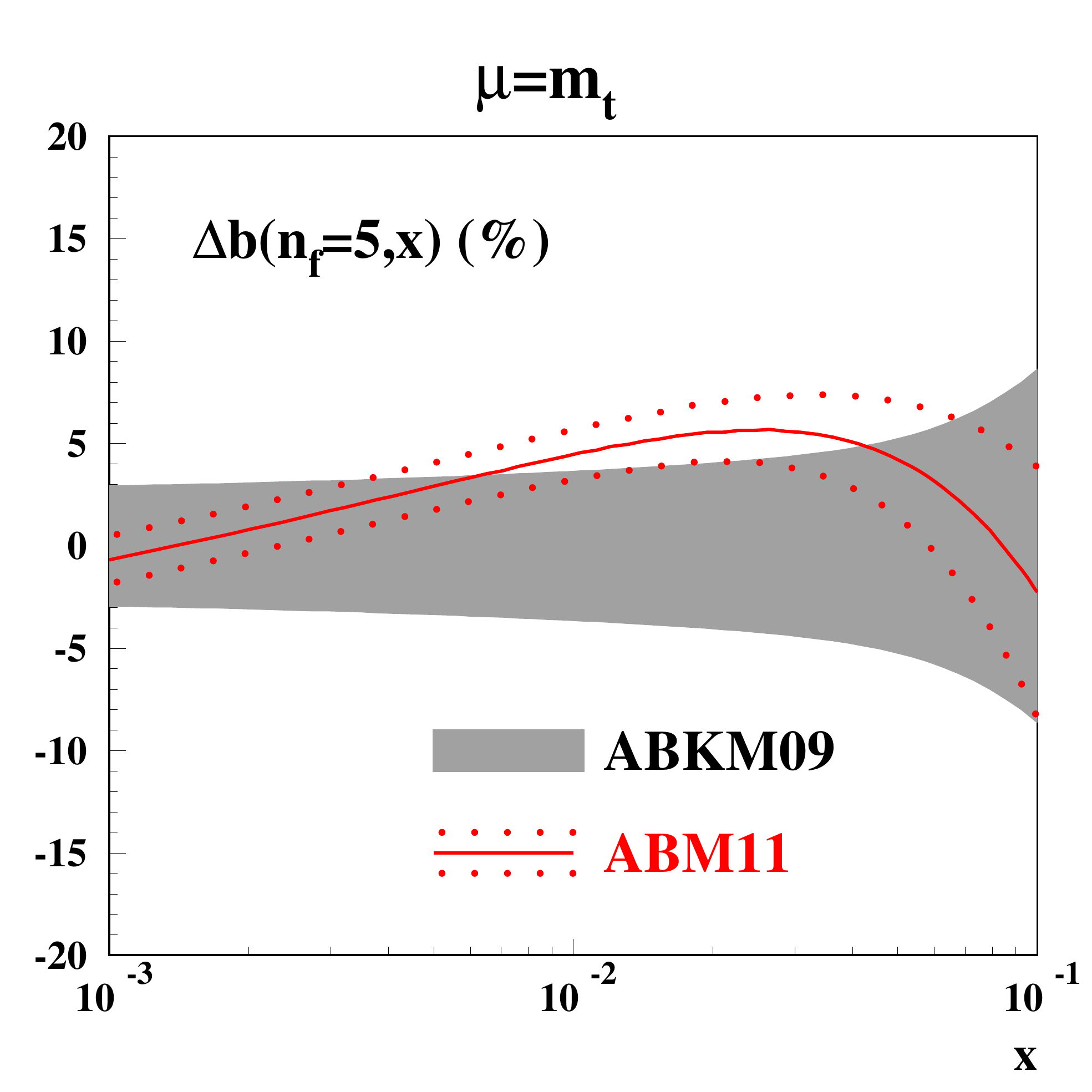}
  \caption{\small
    \label{fig:hq-pdf}
      The charm- (left) and the bottom-quark (right) PDFs obtained in the global fit:
      The dotted (red) lines denote the $\pm 1 \sigma$ band of relative
      uncertainties (in percent) and the solid (red) line indicates the
      central prediction resulting from the fit with
      the running masses~\cite{Alekhin:2010sv}.
      For comparison the shaded (gray) area represents 
     the results of ABKM09~\cite{Alekhin:2009ni}~\cite{Alekhin:2012ig}.
}
\end{figure*}
To a certain extent different VFN scheme prescriptions can be sorted out with 
the data on the semi-inclusive structure function $F_2^{cc}$, which 
correspond to the DIS sub-process with two charmed quarks in the final state.  
Strictly speaking this observable is infrared unsafe due to the 
non-singlet term contributing to perturbative QCD calculations starting from 
the NLO. Rigorous infrared safety of $F_2^{cc}$ can be restored only
with an additional soft cut, e.g. imposed on the $c\bar{c}$ invariant 
mass~\cite{Chuvakin:1999nx}. At the same time the non-singlet 
contribution in $F_2^{cc}$ is numerically small and it can be 
safely disregarded in the analysis of the existing DIS data. 
The recent H1 data on $F_2^{cc}$~\cite{Aaron:2011gp}
 are compared to the predictions 
based on different PDFs in Fig.~\ref{fig:h1cc}.
To provide a comparison consistently we employ the FFN scheme 
with the running-mass definition, 
taking the corresponding 3-flavor PDFs provided by different 
groups and the world-average value of 
$m_c(m_c)=1.27~{\rm GeV}$~\cite{Nakamura:2010zzi}.
This value was also used in our analysis, 
however, the data of Ref.~\cite{Aaron:2011gp}
are not included into our fit. The calculations are performed 
with the open-source code 
OPENQCDRAD~\footnote{\texttt{http://www-zeuthen.desy.de/\~{}alekhin/OPENQCDRAD}}, which contains the 
implementation of the DIS Wilson coefficients up to the NNLO. 
The NNLO ABM11 predictions are in a good agreement with the data. Also 
they agree with the combined H1 and ZEUS data on $F_2^{cc}$, which 
range wider in kinematics~\cite{Lipka}. 
The NLO ABM11 calculations are very close to the NNLO ones
that reflect the improved 
perturbative stability of the running-mass definition scheme. 
The NNLO predictions based on the 
MSTW~\cite{Martin:2009iq} and JR09~\cite{JimenezDelgado:2008hf}
 PDFs are in a good agreement with ours and with the H1 data, while 
the NLO NN21~\cite{Ball:2011mu} predictions systematically undershoot the data.  
The NN21 PDFs are obtained with the FONLL prescription of the VFN 
scheme~\cite{Forte:2010ta}. The FONLL variant of the VFN scheme 
is conceptually similar to the 
prescription of Ref.~\cite{Thorne:2012qh}, however, it allows less flexible
modeling of the massive coefficient functions at $Q^2 \ngg m_h^2$. 
In any case the VFN modeling should reproduce the FFN scheme results at small $Q^2$.
However, the trend demonstrated by the NN21 predictions is opposite and they 
also diverge from the H1 data at small $Q^2$.

The high-$y$ H1 data~\cite{Collaboration:2010ry}
also discriminate different PDFs. The predictions for $F_L$ at low $x$ 
computed with various NLO and NNLO PDFs are compared with the H1 data in 
Fig.~\ref{fig:flh1}. As for the comparison with $F_2^{cc}$ above, we 
employ the FFN scheme with running-mass definition and the 
publicly available 3-flavor PDFs. 
The NNLO ABM11 and JR09 predictions obtained in this way demonstrate good agreement 
with the data in the whole range of $x$, while the MSTW predictions undershoot the data 
at low $x$. This is correlated with the strong fall-off 
of the MSTW gluon distributions at small $x$ taking negative values at 
$x\lesssim 10^{-3}$ at small scales, cf. Fig.~\ref{fig:lowg}.
The observed discrepancy cannot be attributed to the impact of a particular 
choice of the VFN scheme prescription in the MSTW fit
since the low-$x$ tail of the H1 data correspond to small $Q^2$, where the 
VFN scheme employed in the MSTW fit reproduces the FFN one. Due to this  
the small-$x$ MSTW gluon distribution should move up 
and consolidate with the JR09 and ABM ones 
once the H1 data~\cite{Collaboration:2010ry} are included into the MSTW fit. 
The same is also valid for the NN21 PDFs, which overshoot the H1 data at 
small $x$ and undershoot them at larger $x$, cf. Fig.~\ref{fig:lowg}.
\begin{figure*}[t!]
\centerline{
  \includegraphics[scale=0.4]{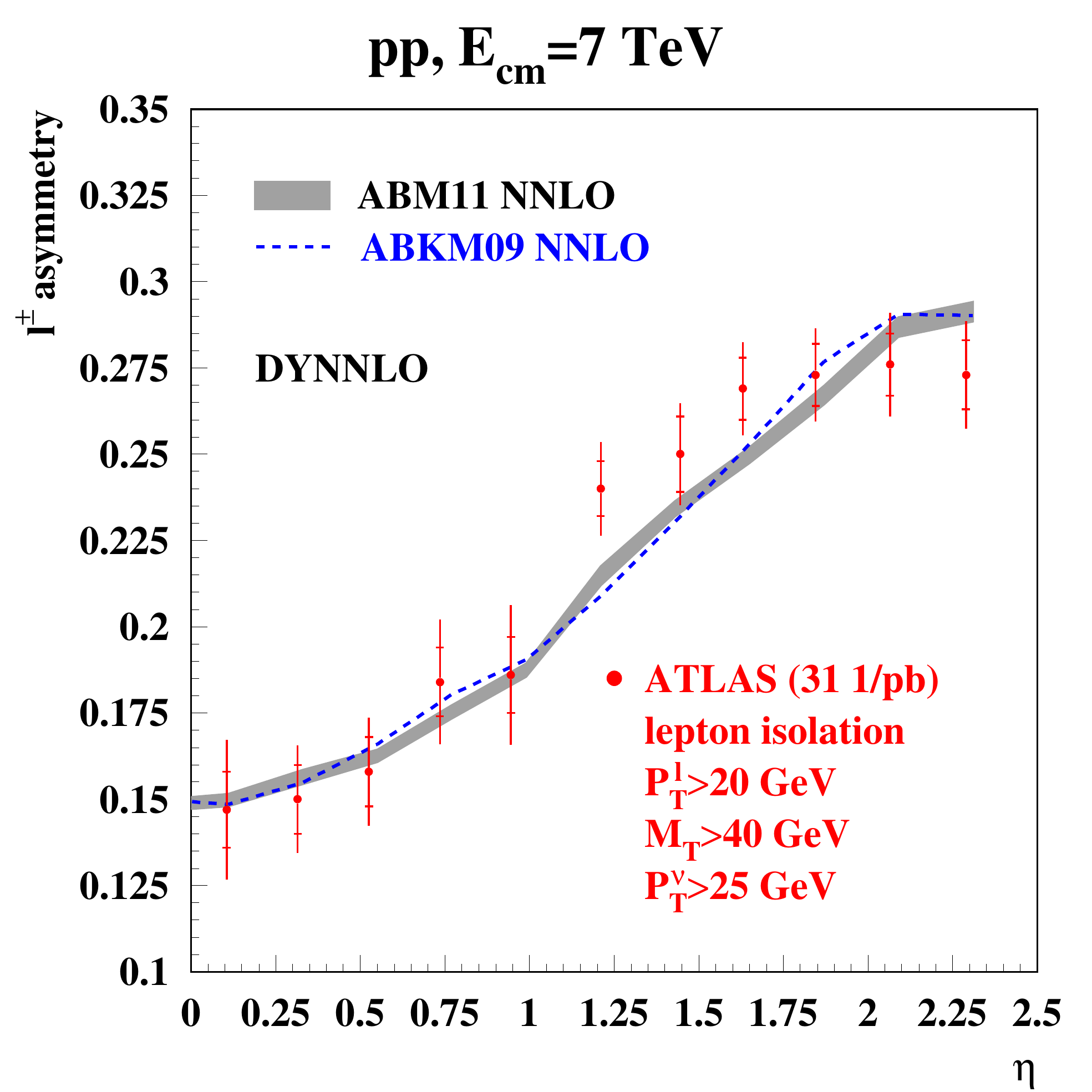}
  \includegraphics[scale=0.4]{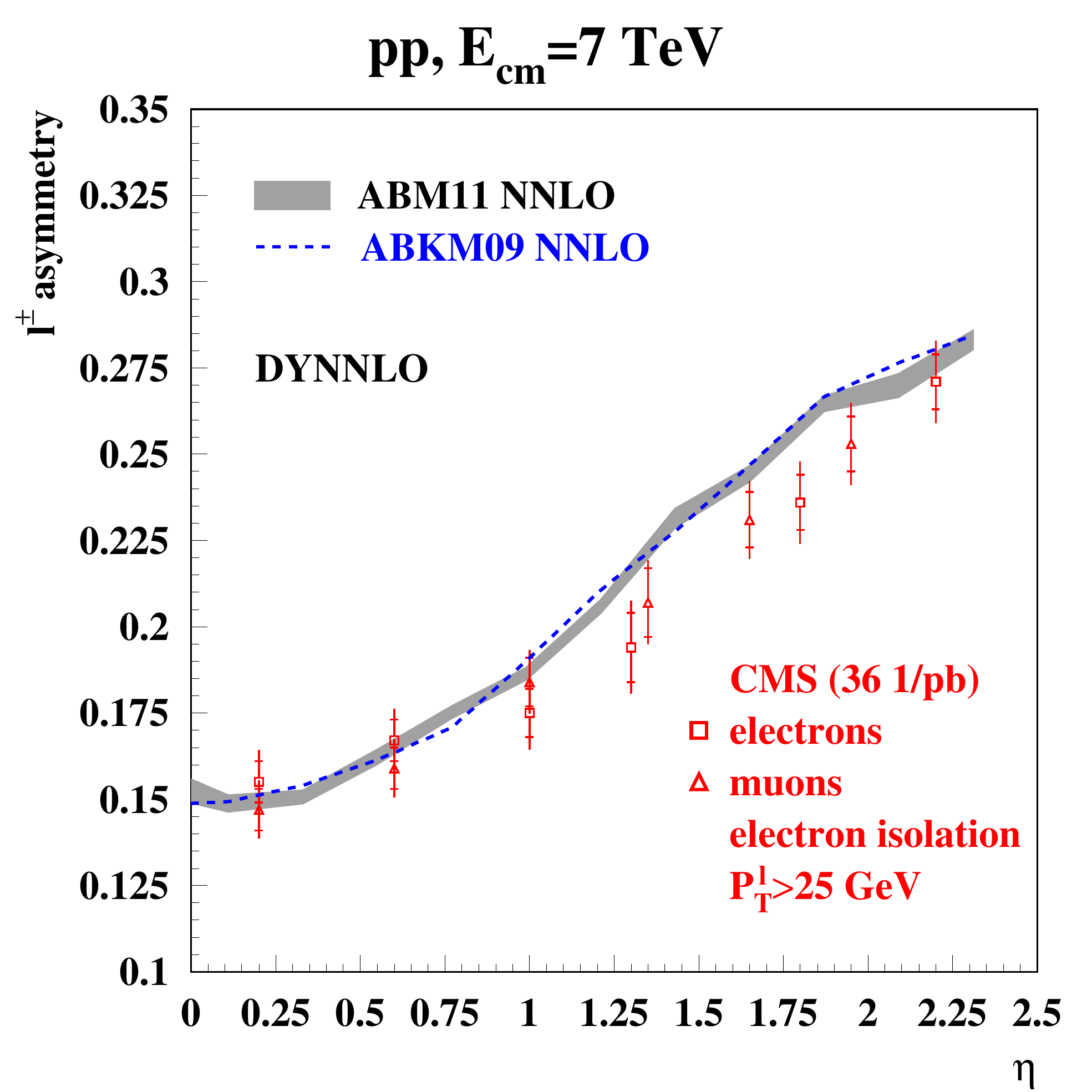}}
  \caption{\small
    \label{fig:lhc}
     The data on charged-lepton asymmetry versus the
     lepton pseudorapidity $\eta$
     obtained by the ATLAS~\cite{Aad:2011yn} (left panel)
and CMS~\cite{Chatrchyan:2011jz} (right panel) experiments
     compared to the NNLO predictions based on the
DYNNLO code of Ref.~\cite{Catani:2010en} and
the ABM11 NNLO PDFs (shaded area showing the integration
uncertainties). The ABKM09 NNLO predictions are given for comparison
by dashes, without the integration uncertainties shown~\cite{Alekhin:2012ig}.
}
\end{figure*}

\begin{figure}[t!]
\centerline{
  \includegraphics[scale=0.4]{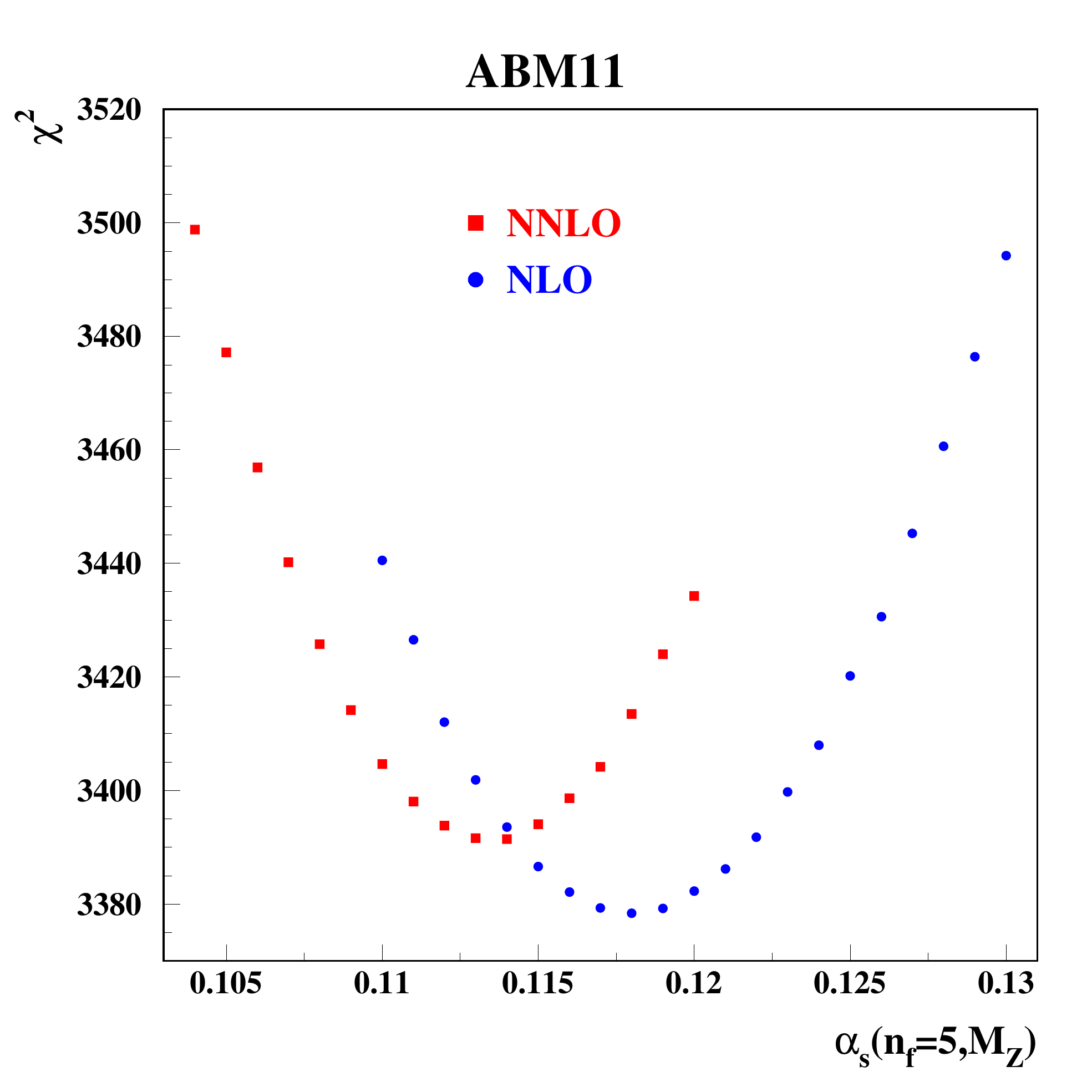}}
  \vspace*{-5mm}
  \caption{\small
    \label{fig:aschi2}
      The $\chi^2$-profile
      as a function of $\alpha_s(M_Z)$ in the present analysis  
      at NLO (circles) and NNLO (squares)~\cite{Alekhin:2012ig}.
  }
\end{figure}

\begin{figure}[th!]
\centerline{
  \includegraphics[scale=0.4]{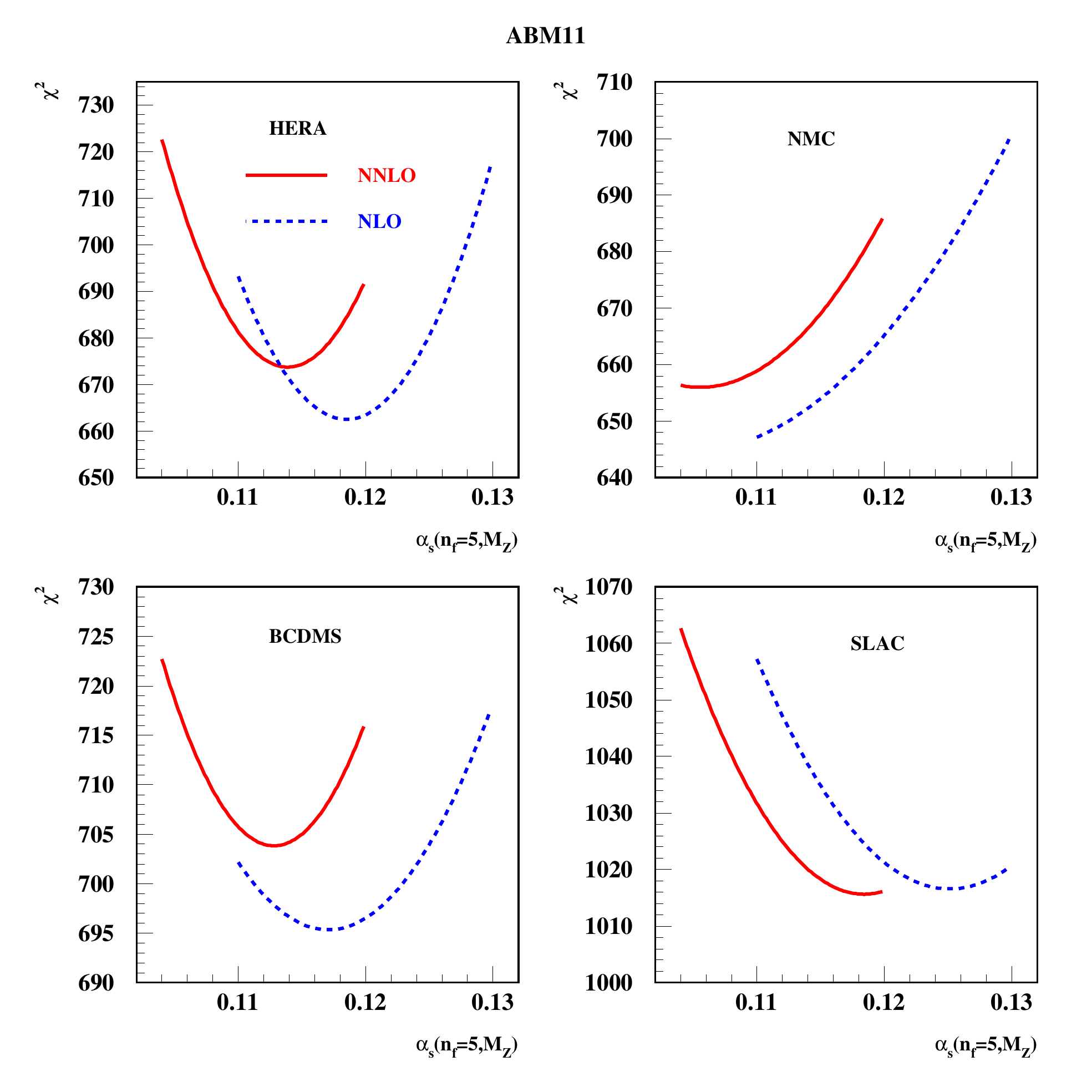}}
  \caption{\small
    \label{fig:scans}
    The $\chi^2$-profile versus the value of $\alpha_s(M_Z)$, 
%    for the data sets of Tab.~\ref{tab:alphas1} 
    for the HERA data~\cite{:2009wt,Collaboration:2010ry},
    the NMC data~\cite{Arneodo:1996qe}, 
    the BCDMS data~\cite{Benvenuti:1989rh,Benvenuti:1989fm}, 
    and the SLAC data~\cite{Whitlow:1990gk,Bodek:1979rx,Atwood:1976ys,Mestayer:1982ba,Gomez:1993ri,Dasu:1993vk}, 
    all obtained in variants of the present analysis with the value of $\alpha_s$ fixed 
    and all other parameters fitted
    (solid lines: NNLO fit, dashes: NLO one)~\cite{Alekhin:2012ig}.
}
\end{figure}

The value of $\alpha_s$ is a necessary ingredient of all QCD calculations
and it should be consistent with the PDFs employed. In our analysis 
we provide this consistency fitting $\alpha_s$
simultaneously with the other parameters of the data model.
In this way we obtain the value of 
\begin{eqnarray}
  \label{eq:annlo}
\alpha_s(M_Z) = 0.1134 \pm 0.0011~~~~({\rm at~NNLO})
\end{eqnarray}
and 
\begin{eqnarray}
  \label{eq:anlo}
\alpha_s(M_Z) = 0.1180 \pm 0.0012~~~~({\rm at~NLO})
\end{eqnarray}
with the uncertainties corresponding to the 
68\%C.L. They are calculated with the standard 
statistical criterion $\Delta \chi^2=1$ taking into account 
correlations of the systematic uncertainties in the 
data within the covariance matrix approach~\cite{Alekhin:2000es}.
Both values in Eqs.~(\ref{eq:annlo},\ref{eq:anlo}) are at variance
due to the $O(0.005)$ scale variation error at NLO.
Once $\alpha_s$ is 
fitted simultaneously with the PDFs and the higher-twist terms in the 
DIS SFs the errors in Eqs.~(\ref{eq:annlo},\ref{eq:anlo}) take into account  
uncertainties in the latter as well. The ABM11 value of $\alpha_s$ 
is in a very good agreement with  
our earlier ABKM09 value and the error is somewhat 
smaller due to the improved  accuracy of the HERA data.
The profiles of $\chi^2$ versus $\alpha_s$,
obtained in the NLO and NNLO 
variants of the ABM11 analysis with the value of $\alpha_s$ fixed 
and all other parameters fitted, are nearly parabolic at the minimum 
with the shapes determined by Eqs.~(\ref{eq:annlo},\ref{eq:anlo}), 
cf. Fig.~\ref{fig:aschi2}. 
This gives an additional justifications of using the standard 
statistical criteria to estimate the parameter uncertainties in our fit. 
The $\chi^2$ profiles for the separate data sets employed in the fit  
are also nearly parabolic, cf. Fig.~\ref{fig:scans}. 
The value of $\alpha_s$ preferred by the HERA and 
BCDMS data are in a good agreement, while the NMC and SLAC data 
prefer somewhat smaller and bigger value, respectively. 
The SLAC and the NMC data are sensitive to the higher-twist 
contribution, cf. discussion in~\cite{Alekhin:2012ig,Alekhin:2011ey}. 
Note, the cut on the hadronic invariant mass $W$, which is commonly 
used in the global PDF fits, does not allow to get rid of the 
impact of the higher-twist terms on $\alpha_s$. Indeed, 
in a variant of our fit with the cut of $W^2>12.5~{\rm GeV}^2$
imposed and all higher-twist terms fixed at 0
we obtain $\alpha_s(M_Z) = 0.1191 \pm 0.0006$ at NNLO. This 
is substantially bigger than the value of $\alpha_s$ in
Eq.~(\ref{eq:annlo}). It is worth noting that in this way 
we approach the values of $\alpha_s$ obtained in the  
PDF fits~\cite{Martin:2009iq,Ball:2011us}, performed with  
no higher-twist terms are taken into account. The value of 
$\alpha_s$ is also sensitive to the treatment of the 
correlated uncertainties in the data. E.g. in the NNLO MSTW 
fit~\cite{Martin:2009iq}  
the HERA and NMC data prefer value of $\alpha_s(M_Z) \gtrsim 0.12$, 
contrary to our findings, cf. Fig.~\ref{fig:scans}. Note 
that in~\cite{Martin:2009iq}
the NMC and HERA systematics errors are combineid in 
quadrature. To study impact of this approximation we 
performed a trial NNLO ABKM09 fit with the same treatment of the 
NMC and HERA systematics and found that the value of $\alpha_s(M_Z)$
shifted up by +0.0029 as compared to the nominal ABKM value. 
Many other aspects of 
our analysis, which may affect the value of $\alpha_s$,
are also different from~\cite{Martin:2009iq,Ball:2011us}: 
basic relations for the DIS cross sections, data normalization, 
etc, cf. Ref.~\cite{Alekhin:2012ig} for a detailed discussion. 
These differences make a detailed comparison of our results 
with~\cite{Martin:2009iq,Ball:2011us} difficult.

While the 3-flavor FFN scheme is nicely sufficient for description of the 
DIS data,  
the 5-flavor factorization schemes is most often justified 
for the collider phenomenology in view of much bigger scales involved.  
In cases the 4-flavor scheme may be also relevant.
The 4(5)-flavor ABM11 PDFs are matched with the 3(4)-flavor ones at 
the scale of $m_c$ and $m_b$, respectively, employing massive OMEs 
taken in the running-mass definition~\cite{Alekhin:2010sv}.  
The 4- and 5-flavor PDFs at larger scales are generated
by means of the massless QCD evolution with these boundary conditions. 
The heavy-quark PDFs are particularly sensitive 
to the values of $m_{c},m_{b}$. 
Taking advantage of the \MSbar definition we fix them at 
the PDG values of Eqs.(\ref{eq:mcmbinp},\ref{eq:mbmbinp}). 
Due to the change in the heavy flavor treatment and impact of the new data 
included into the fit, the ABM11 heavy-quark PDFs differ from the ABKM09 
ones, cf. Fig.~\ref{fig:hq-pdf}. The uncertainties in the 
heavy-quark PDFs are to a large extend defined by the uncertainties 
in $m_{c},m_{b}$. In our fit the latter are calculated as a sensitivity 
of the fitted data supplemented by the PDG constraints of 
Eqs.(\ref{eq:mcmbinp},\ref{eq:mbmbinp}) to  $m_{c},m_{b}$. The 
uncertainty in $m_{b}$ obtained in this way coincides with one 
in Eq.(\ref{eq:mcmbinp}), while for $m_{c}$ it reduces to 0.06~GeV due to 
impact of the inclusive HERA data. The uncertainties in the 
heavy-quark PDFs estimated with these constraints on $m_{c},m_{b}$ 
are essentially reduced as compared to the ABKM09 case. This improvement 
is especially important for the hadronic single-top production
driven by the initial-state $b$-quarks and for the 
Higgs production through the vector-boson-fusion channel, which is sensitive to 
the $c$-quark distribution~\cite{Bolzoni:2010xr}.


The hadronic jet production provides an additional constraint on the 
PDFs, in particular on the large-$x$ gluon 
distribution~\cite{Martin:2009iq,Lai:2010vv,Ball:2011mu}. However, 
the calculation of the full NNLO QCD corrections to this process is still in
progress (see~\cite{GehrmannDeRidder:2011aa,Bolzoni:2010bt} and references therein). 
This precludes a consistent use of the Tevatron jet data in our NNLO PDF fit. 
Nevertheless, in order to check any potential impact of the jet Tevatron data on our PDFs we 
have performed trial variants of the NNLO ABKM09 fit with the Tevatron jet data added~\cite{Alekhin:2011cf}.
The NLO QCD corrections~\cite{Nagy:2001fj,Nagy:2003tz} and the partial (soft gluon enhanced) 
NNLO corrections due to threshold resummation~\cite{Kidonakis:2000gi} have been computed with the {\tt FastNLO}
tool~\cite{Kluge:2006xs,Wobisch:2011ij}.
\begin{table}[h!]\centering
\begin{tabular}{|l|c|c|}
\hline
\multicolumn{1}{|c|}{Experiment} &
\multicolumn{2}{c|}{$\alpha_s(M_Z)$} \\
\cline{2-3}
\multicolumn{1}{|c|}{ } &
\multicolumn{1}{c|}{NLO} &
\multicolumn{1}{c|}{NNLO$^*$} \\
\hline 
D0  1 jet             & 
                       $0.1190 \pm 0.0011$ & $0.1149(12)$ \\
D0  2 jet                   & $0.1174(9)$ & $0.1145(9)$ \\   
CDF 1 jet (cone)            & $0.1181(9)9$ & $0.1134(9)$ \\   
CDF 1 jet ($k_\perp$)       & $0.1181(10)$ & $0.1143(9)$ \\   
\hline
ABM11                       & $0.1180(12)$ & $0.1134(11)$ \\   
\hline 
\end{tabular} 
\renewcommand{\arraystretch}{1}  
\caption{ \small
\label{tab:alphas2}
The values of $\alpha_s({M_Z})$ based on including individual 
data sets of Tevatron jet data~\cite{Abulencia:2007ez,Aaltonen:2008eq,Abazov:2008hua,Abazov:2010fr} 
into the analysis at NLO. 
The NNLO$^*$ fit refers to the NNLO analysis of the DIS and DY data together with 
the NLO and soft gluon resummation corrections (next-to-leading logarithmic accuracy) 
for the 1 jet inclusive data, cf. \cite{Kidonakis:2000gi,Alekhin:2011cf}.}
\end{table}

\begin{figure}[th]
\centering
    {
    \includegraphics[width=8cm,height=9.5cm]{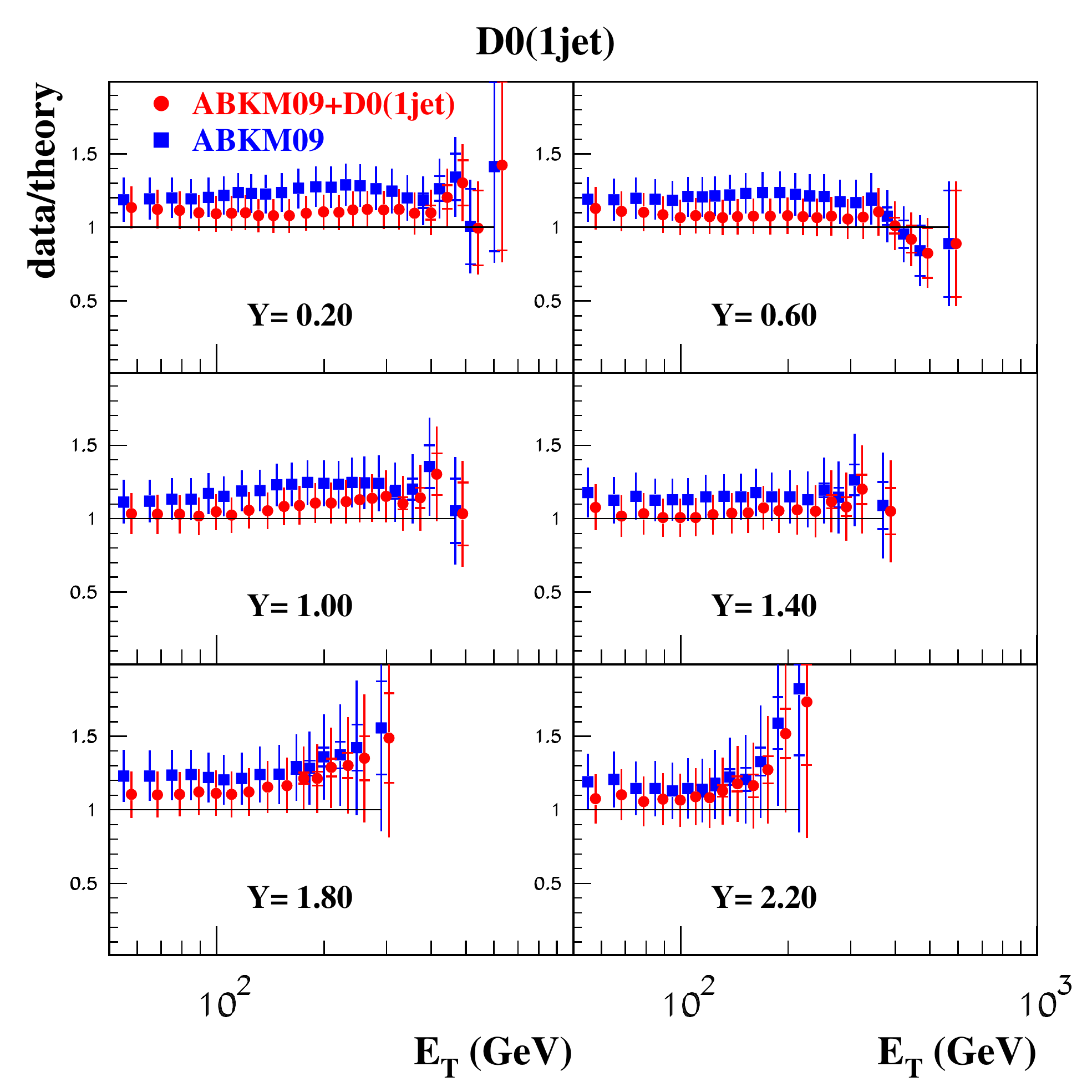}
    }
    \caption{\small
      \label{fig:jets1}
      Cross section data for 1-jet inclusive production from the D0 collaboration~\cite{Abazov:2008hua}
      as a function of the jet's transverse energy $E_T$  
      for the renormalization and factorization scales equal to $E_T$
      compared to the result of~\cite{Alekhin:2009ni} (circles) 
      and a re-fit including this data (squares) including 
      the NNLO threshold resummation corrections to the jet production~\cite{Kidonakis:2000gi}. 
    }
\end{figure}

\begin{figure}[ht!]
\centering
    {
    \includegraphics[width=8cm,height=9.5cm]{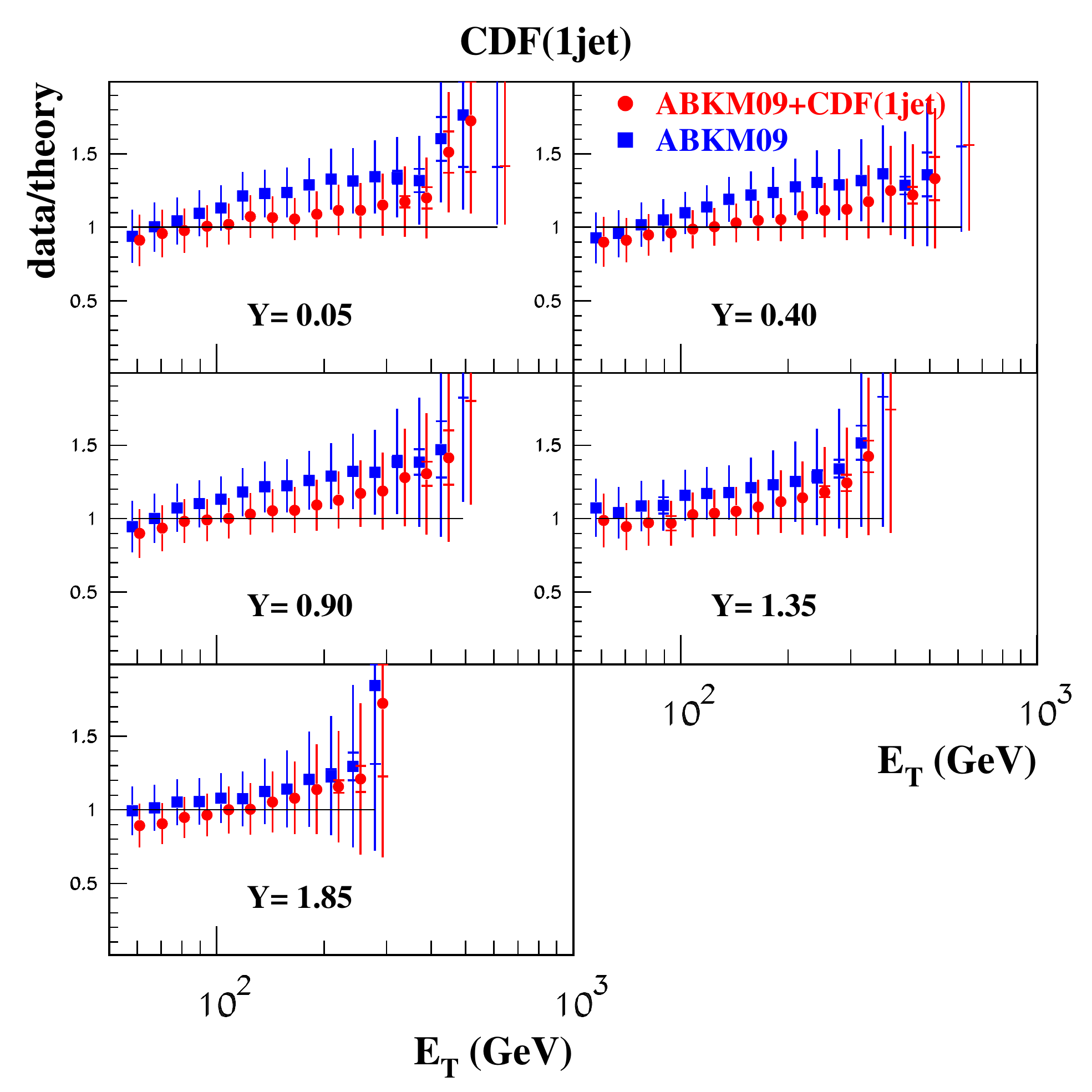}
    }
    \caption{\small
      \label{fig:jets2}
      Same as Fig.~\ref{fig:jets1} 
      for the cross section data for 1-jet inclusive production from 
      the CDF collaboration
      using a $k_T$ jet algorithm~\cite{Abulencia:2007ez}.
    }
\end{figure}

In general, the Tevatron jet data overshoot the ABKM09 predictions, nevertheless they can be smoothly accommodated in the fit. 
The typical value of $\chi^2/NDP\approx 1$ is achieved with account of the error correlations for the jet data sets 
of~\cite{Abazov:2008hua,Abazov:2010fr,Abulencia:2007ez,Aaltonen:2008eq} 
once they are included into the NNLO ABKM09 fit. 
Meanwhile the various data sets demonstrate a somewhat different trend with respect to the ABKM09 predictions. 
E.g., the off-set of the D0 inclusive jet data~\cite{Abazov:2008hua} does not depend on the jet energy $E_T$ 
and therefore may be attributed to the impact of the currently missing full
NNLO corrections, cf. Figs.~\ref{fig:jets1} and \ref{fig:jets2}.
In contrast, for the CDF data of~\cite{Abulencia:2007ez} obtained with the $k_T$ jet algorithm 
the pulls rise with $E_T$ and can be reduced only by means of a modification of the PDF shapes.

\begin{table*}[ht!]
\centering
\begin{tabular}{|c|c|c|c|c|c|}
\hline
  $\sigma(H) [pb]$
  & ABKM09 
  & D0 1-jet inc. 
  & D0 di-jet 
  & CDF 1-jet inc.
  & CDF 1-jet inc.
\\
  & 
  & 
  & 
  & (cone)
  & ($k_T$)
\\[0.5ex]
\hline
% mh = 115
%Tevatron(1.96) & 
%    {\bf 0.885(55)} & 0.981(33) &  0.954(30) & 0.932(28)  & 0.962(28) 
%\\[0.5ex]
%LHC(7)
%    & 
%    {\bf 15.72(45)} & 16.08(32) &  16.10(29) & 15.45(30) & 15.81(30) 
%\\
% mh = 120
Tevatron(1.96) & 
    {\bf 0.770(50)} & 0.859(29) &  0.833(27) & 0.815(25)  & 0.842(25) 
\\
LHC(7)
    & 
    {\bf 14.34(41)} & 14.68(29) &  14.69(27) & 14.11(28) & 14.44(27) 
\\
\hline
\end{tabular}
\caption{\small
\label{tab:hxsvalues}
The predicted cross sections for Higgs boson production in ggF with 
$m_H = 120$~GeV at Tevatron ($\sqrt{s}=1.96$~TeV) and at LHC ($\sqrt{s}=7$~TeV) from 
NNLO variants of the ABKM09 fit~\cite{Alekhin:2009ni} corresponding to Tab.~\ref{tab:alphas2}.
The uncertainty in brackets refers to the $1\sigma$ standard deviation for the
combined uncertainty on the PDFs and the value of $\alpha_s(M_Z)$.
The values in bold correspond to the published result~\cite{Alekhin:2010dd}.
}
\end{table*}

The values of $\alpha_s$ extracted from the trial ABKM09 fits with the Tevatron jet data included 
are compared with the nominal ABKM09 value in Tab.~\ref{tab:alphas2}.
At most, they are bigger by 1$\sigma$, while for the CDF cone jet algorithm data~\cite{Aaltonen:2008eq} 
the central value of $\alpha_s$ is even the same. 
The predictions for the light Higgs production cross section, 
which are defined by the gluon distribution at $x\lesssim 0.1$, are also not very 
sensitive to the constraints coming from the Tevatron data, cf. Tab.~\ref{tab:hxsvalues}.
The impact of the Tevatron jet data on the large-$x$ gluon distribution is more significant. 
However, in this context we note that the trend of the first LHC data on the jet production 
with respect to the various PDF predictions is different from the Tevatron measurements. 
The ABKM09 predictions are in better agreement with the CMS and ATLAS inclusive data 
of~\cite{Rabbertz:1368241,Aad:2011fc} than the predictions based on the PDFs of~\cite{Martin:2009iq,Lai:2010vv,Ball:2011mu}, 
which were tuned to the Tevatron inclusive jet data. 
Jet data from LHC is still subject to large systematic errors, though.
Note also that the Tevatron dijet and 3-jet production data are in good 
agreement with the ABKM09 predictions~\cite{Wobisch:2012iu}, in contrast to the  
case of inclusive jet production at Tevatron. 
These ambiguities in the data as well as the limitations in the current theoretical treatment   
prevent the use of hadronic jet data in our fit.

The data on $W/Z$ productions being produced by the LHC
experiments also may help to improve the PDF accuracy.
The charged-lepton asymmetry data~\cite{Aad:2011yn,Chatrchyan:2011jz} 
obtained by the ATLAS and CMS experiments are compared 
to the NNLO predictions based on the ABM11 PDFs in Fig.~\ref{fig:lhc}.  
All differential distributions for $W$- and $Z$-boson production are
computed with the fully exclusive NNLO program 
{\tt DYNNLO}~\cite{Catani:2010en}, 
which allows to take into account the kinematical cuts imposed in the experiments (cf. Fig.~\ref{fig:lhc}). 
The overall agreement with both experiments is sufficiently good. At  
values of $\eta\sim 1.5$ for the lepton pseudo-rapidity $\eta$ the data show 
a different trend with respect to the predictions, however 
the discrepancy is within the data uncertainties. 
Preliminary data on the charge-lepton asymmetry at large
rapidities obtained by the LHCb collaboration~\cite{Amhis:2012gj} are also in 
good agreement with the ABKM09 predictions. 
 To check the impact of the 
LHC charged-lepton asymmetry data on our fit, 
we have performed a variant of the ABM11 analysis which consists of 
adding the data of~\cite{Aad:2011yn,Chatrchyan:2011jz} to the fit. 
We have found that the impact of those data is only marginal 
in view of still big experimental uncertainties.


{\bf Summary.}
We have discussed different schemes for the treatment of the heavy flavor component in DIS. 
In the kinematic region $Q^2 \simeq m_h^2$ all schemes need to match 
to the 3-flavor scheme, which allows for a consistent comparison of the various PDF sets. 
We have performed this benchmark exercise with the help to the OPENQCDRAD code 
for standardized precision comparison.
Within the framework of the ABM11 PDF analysis, which uses the running mass scheme for
the heavy quarks, we find that the FFN is completely sufficient for describing
the existing DIS data. 

We have also given detailed information on theoretical and experimental
improvements which are of relevance for the low-$x$ PDFs, especially for the
gluon and we have briefly sketched the implications for LHC phenomenology, so
that differences of ABM11 with respect to other PDF sets can be explained. 
More benchmark comparisons with the help of OPENQCDRAD 
can be found elsewhere~\cite{Alekhin:2012ig}
and are expected in the future.

{\bf Acknowledgments.}
This work has been supported by Helmholtz Gemeinschaft under contract VH-HA-101 ({\it Alliance Physics at the Terascale}),
by the Deutsche Forschungsgemeinschaft in Sonderforschungs\-be\-reich/Transregio~9
and by the European Commission through contract PITN-GA-2010-264564 ({\it LHCPhenoNet}).

\bibliographystyle{elsarticle-num}
\bibliography{alekhin.bib}

\begin{thebibliography}{10}
\expandafter\ifx\csname url\endcsname\relax
  \def\url#1{\texttt{#1}}\fi
\expandafter\ifx\csname urlprefix\endcsname\relax\def\urlprefix{URL }\fi
\expandafter\ifx\csname href\endcsname\relax
  \def\href#1#2{#2} \def\path#1{#1}\fi

\bibitem{Witten:1975bh}
E.~Witten, {Heavy Quark Contributions to Deep Inelastic Scattering}, Nucl.Phys.
  B104 (1976) 445--476.
\newblock \href {http://dx.doi.org/10.1016/0550-3213(76)90111-5}
  {\path{doi:10.1016/0550-3213(76)90111-5}}.

\bibitem{Laenen:1992zk}
E.~Laenen, S.~Riemersma, J.~Smith, W.~van Neerven, {Complete $O(\alpha_s)$
  corrections to heavy flavor structure functions in electroproduction},
  Nucl.Phys. B392 (1993) 162--228.
\newblock \href {http://dx.doi.org/10.1016/0550-3213(93)90201-Y}
  {\path{doi:10.1016/0550-3213(93)90201-Y}}.

\bibitem{Gottschalk:1980rv}
T.~Gottschalk, {Chromodynamic corrections to neutrino production of heavy
  quarks}, Phys.Rev. D23 (1981) 56.
\newblock \href {http://dx.doi.org/10.1103/PhysRevD.23.56}
  {\path{doi:10.1103/PhysRevD.23.56}}.

\bibitem{Gluck:1996ve}
M.~Gl{\"u}ck, S.~Kretzer, E.~Reya, {The Strange sea density and charm
  production in deep inelastic charged current processes}, Phys.Lett. B380
  (1996) 171--176.
\newblock \href {http://arxiv.org/abs/hep-ph/9603304}
  {\path{arXiv:hep-ph/9603304}}, \href
  {http://dx.doi.org/10.1016/0370-2693(96)00456-X,
  10.1016/0370-2693(96)00456-X} {\path{doi:10.1016/0370-2693(96)00456-X,
  10.1016/0370-2693(96)00456-X}}.

\bibitem{Laenen:1998kp}
E.~Laenen, S.-O. Moch, {Soft gluon resummation for heavy quark
  electroproduction}, Phys.Rev. D59 (1999) 034027.
\newblock \href {http://arxiv.org/abs/hep-ph/9809550}
  {\path{arXiv:hep-ph/9809550}}, \href
  {http://dx.doi.org/10.1103/PhysRevD.59.034027}
  {\path{doi:10.1103/PhysRevD.59.034027}}.

\bibitem{Corcella:2003ib}
G.~Corcella, A.~D. Mitov, {Soft gluon resummation for heavy quark production in
  charged current deep inelastic scattering}, Nucl.Phys. B676 (2004) 346--364.
\newblock \href {http://arxiv.org/abs/hep-ph/0308105}
  {\path{arXiv:hep-ph/0308105}}, \href
  {http://dx.doi.org/10.1016/j.nuclphysb.2003.10.027}
  {\path{doi:10.1016/j.nuclphysb.2003.10.027}}.

\bibitem{Alekhin:2008hc}
S.~Alekhin, S.~Moch, {Higher order QCD corrections to charged-lepton
  deep-inelastic scattering and global fits of parton distributions},
  Phys.Lett. B672 (2009) 166--171.
\newblock \href {http://arxiv.org/abs/0811.1412} {\path{arXiv:0811.1412}},
  \href {http://dx.doi.org/10.1016/j.physletb.2009.01.004}
  {\path{doi:10.1016/j.physletb.2009.01.004}}.

\bibitem{Presti:2010pd}
N.~Lo~Presti, H.~Kawamura, S.~Moch, A.~Vogt, {Threshold-improved predictions
  for charm production in deep-inelastic scattering}, PoS DIS2010 (2010) 163.
\newblock \href {http://arxiv.org/abs/1008.0951} {\path{arXiv:1008.0951}}.

\bibitem{Alekhin:2010sv}
S.~Alekhin, S.~Moch, {Heavy-quark deep-inelastic scattering with a running
  mass}, Phys.Lett. B699 (2011) 345--353.
\newblock \href {http://arxiv.org/abs/1011.5790} {\path{arXiv:1011.5790}},
  \href {http://dx.doi.org/10.1016/j.physletb.2011.04.026}
  {\path{doi:10.1016/j.physletb.2011.04.026}}.

\bibitem{:2009wt}
F.~Aaron, et~al., {Combined Measurement and QCD Analysis of the Inclusive
  $e^{\pm}$p Scattering Cross Sections at HERA}, JHEP 1001 (2010) 109.
\newblock \href {http://arxiv.org/abs/0911.0884} {\path{arXiv:0911.0884}},
  \href {http://dx.doi.org/10.1007/JHEP01(2010)109}
  {\path{doi:10.1007/JHEP01(2010)109}}.

\bibitem{Nakamura:2010zzi}
K.~Nakamura, et~al., {Review of particle physics}, J.Phys.G G37 (2010) 075021.
\newblock \href {http://dx.doi.org/10.1088/0954-3899/37/7A/075021}
  {\path{doi:10.1088/0954-3899/37/7A/075021}}.

\bibitem{Alekhin:2009ni}
S.~Alekhin, J.~Bl{\"u}mlein, S.~Klein, S.~Moch, {The 3, 4, and 5-flavor NNLO
  Parton from Deep-Inelastic-Scattering Data and at Hadron Colliders},
  Phys.Rev. D81 (2010) 014032.
\newblock \href {http://arxiv.org/abs/0908.2766} {\path{arXiv:0908.2766}},
  \href {http://dx.doi.org/10.1103/PhysRevD.81.014032}
  {\path{doi:10.1103/PhysRevD.81.014032}}.

\bibitem{Adloff:2000qk}
C.~Adloff, et~al., {Deep inelastic inclusive e p scattering at low x and a
  determination of $\alpha_s$}, Eur.Phys.J. C21 (2001) 33--61.
\newblock \href {http://arxiv.org/abs/hep-ex/0012053}
  {\path{arXiv:hep-ex/0012053}}, \href
  {http://dx.doi.org/10.1007/s100520100720} {\path{doi:10.1007/s100520100720}}.

\bibitem{Chekanov:2001qu}
S.~Chekanov, et~al., {Measurement of the neutral current cross-section and
  $F_2$ structure function for deep inelastic $e^{+}$ p scattering at HERA},
  Eur.Phys.J. C21 (2001) 443--471.
\newblock \href {http://arxiv.org/abs/hep-ex/0105090}
  {\path{arXiv:hep-ex/0105090}}, \href
  {http://dx.doi.org/10.1007/s100520100749} {\path{doi:10.1007/s100520100749}}.

\bibitem{Alekhin:2008mb}
S.~Alekhin, S.~A. Kulagin, R.~Petti, {Determination of Strange Sea
  Distributions from Neutrino-Nucleon Deep Inelastic Scattering}, Phys.Lett.
  B675 (2009) 433--440.
\newblock \href {http://arxiv.org/abs/0812.4448} {\path{arXiv:0812.4448}},
  \href {http://dx.doi.org/10.1016/j.physletb.2009.04.033}
  {\path{doi:10.1016/j.physletb.2009.04.033}}.

\bibitem{Boer:2011fh}
D.~Boer, M.~Diehl, R.~Milner, R.~Venugopalan, W.~Vogelsang, et~al., {Gluons and
  the quark sea at high energies: Distributions, polarization,
  tomography,~}\href {http://arxiv.org/abs/1108.1713} {\path{arXiv:1108.1713}}.

\bibitem{Collaboration:2010ry}
F.~Aaron, C.~Alexa, V.~Andreev, S.~Backovic, A.~Baghdasaryan, et~al.,
  {Measurement of the Inclusive $e^{\pm}p$ Scattering Cross Section at High
  Inelasticity y and of the Structure Function FL}, Eur.Phys.J. C71 (2011)
  1579.
\newblock \href {http://arxiv.org/abs/1012.4355} {\path{arXiv:1012.4355}},
  \href {http://dx.doi.org/10.1140/epjc/s10052-011-1579-4}
  {\path{doi:10.1140/epjc/s10052-011-1579-4}}.

\bibitem{Alekhin:2012ig}
S.~Alekhin, J.~Bl{\"u}mlein, S.~Moch, {Parton distribution functions and
  benchmark cross sections at NNLO,~}\href {http://arxiv.org/abs/1202.2281}
  {\path{arXiv:1202.2281}}.

\bibitem{Ball:2011mu}
R.~D. Ball, V.~Bertone, F.~Cerutti, L.~Del~Debbio, S.~Forte, et~al., {Impact of
  Heavy Quark Masses on Parton Distributions and LHC Phenomenology}, Nucl.Phys.
  B849 (2011) 296--363.
\newblock \href {http://arxiv.org/abs/1101.1300} {\path{arXiv:1101.1300}},
  \href {http://dx.doi.org/10.1016/j.nuclphysb.2011.03.021}
  {\path{doi:10.1016/j.nuclphysb.2011.03.021}}.

\bibitem{Martin:2009iq}
A.~Martin, W.~Stirling, R.~Thorne, G.~Watt, {Parton distributions for the LHC},
  Eur.Phys.J. C63 (2009) 189--285.
\newblock \href {http://arxiv.org/abs/0901.0002} {\path{arXiv:0901.0002}},
  \href {http://dx.doi.org/10.1140/epjc/s10052-009-1072-5}
  {\path{doi:10.1140/epjc/s10052-009-1072-5}}.

\bibitem{JimenezDelgado:2008hf}
P.~Jimenez-Delgado, E.~Reya, {Dynamical NNLO parton distributions}, Phys.Rev.
  D79 (2009) 074023.
\newblock \href {http://arxiv.org/abs/0810.4274} {\path{arXiv:0810.4274}},
  \href {http://dx.doi.org/10.1103/PhysRevD.79.074023}
  {\path{doi:10.1103/PhysRevD.79.074023}}.

\bibitem{Aaron:2011gp}
F.~Aaron, et~al., {Measurement of $D^{\pm}$ Meson Production and determination
  of $F_2^{c\bar{c}}$ at low $Q^2$ in Deep-Inelastic Scattering at HERA},
  Eur.Phys.J. C71 (2011) 1769.
\newblock \href {http://arxiv.org/abs/1106.1028} {\path{arXiv:1106.1028}},
  \href {http://dx.doi.org/10.1140/epjc/s10052-011-1769-0}
  {\path{doi:10.1140/epjc/s10052-011-1769-0}}.

\bibitem{Harris:1995tu}
B.~Harris, J.~Smith, {Heavy quark correlations in deep inelastic
  electroproduction}, Nucl.Phys. B452 (1995) 109--160.
\newblock \href {http://arxiv.org/abs/hep-ph/9503484}
  {\path{arXiv:hep-ph/9503484}}, \href
  {http://dx.doi.org/10.1016/0550-3213(95)00256-R}
  {\path{doi:10.1016/0550-3213(95)00256-R}}.

\bibitem{Arneodo:1996qe}
M.~Arneodo, et~al., {Measurement of the proton and deuteron structure
  functions, $F_2$(p) and $F_2$(d), and of the ratio $\sigma_L/\sigma_T$},
  Nucl.Phys. B483 (1997) 3--43.
\newblock \href {http://arxiv.org/abs/hep-ph/9610231}
  {\path{arXiv:hep-ph/9610231}}, \href
  {http://dx.doi.org/10.1016/S0550-3213(96)00538-X}
  {\path{doi:10.1016/S0550-3213(96)00538-X}}.

\bibitem{Benvenuti:1989rh}
A.~Benvenuti, et~al., {A High Statistics Measurement of the Proton Structure
  Functions $F_2 (x, Q^2)$ and R from Deep Inelastic Muon Scattering at High
  $Q^2$}, Phys.Lett. B223 (1989) 485.
\newblock \href {http://dx.doi.org/10.1016/0370-2693(89)91637-7}
  {\path{doi:10.1016/0370-2693(89)91637-7}}.

\bibitem{Benvenuti:1989fm}
A.~Benvenuti, et~al., {A high statistics measurement of the deuteron structure
  functions $F_2 (x, Q^2)$ and R from deep inelastic muon scattering at high
  $Q^2$}, Phys.Lett. B237 (1990) 592.
\newblock \href {http://dx.doi.org/10.1016/0370-2693(90)91231-Y}
  {\path{doi:10.1016/0370-2693(90)91231-Y}}.

\bibitem{Whitlow:1990gk}
L.~Whitlow, S.~Rock, A.~Bodek, E.~Riordan, S.~Dasu, {A Precise extraction of $R
  = \sigma_L / \sigma_T$ from a global analysis of the SLAC deep inelastic e p
  and e d scattering cross-sections}, Phys.Lett. B250 (1990) 193--198.
\newblock \href {http://dx.doi.org/10.1016/0370-2693(90)91176-C}
  {\path{doi:10.1016/0370-2693(90)91176-C}}.

\bibitem{Bodek:1979rx}
A.~Bodek, M.~Breidenbach, D.~Dubin, J.~Elias, J.~I. Friedman, et~al.,
  {Experimental Studies of the Neutron and Proton Electromagnetic Structure
  Functions}, Phys.Rev. D20 (1979) 1471--1552.
\newblock \href {http://dx.doi.org/10.1103/PhysRevD.20.1471}
  {\path{doi:10.1103/PhysRevD.20.1471}}.

\bibitem{Atwood:1976ys}
W.~Atwood, E.~D. Bloom, R.~Cottrell, H.~DeStaebler, M.~Mestayer, et~al.,
  {Inelastic electron Scattering from Hydrogen at 50-Degrees and 60-Degrees},
  Phys.Lett. B64 (1976) 479.
\newblock \href {http://dx.doi.org/10.1016/0370-2693(76)90127-1}
  {\path{doi:10.1016/0370-2693(76)90127-1}}.

\bibitem{Mestayer:1982ba}
M.~Mestayer, W.~Atwood, E.~D. Bloom, R.~Cottrell, H.~DeStaebler, et~al., {The
  ratio $\sigma_L / \sigma_T$ from deep inelastic electron scattering},
  Phys.Rev. D27 (1983) 285.
\newblock \href {http://dx.doi.org/10.1103/PhysRevD.27.285}
  {\path{doi:10.1103/PhysRevD.27.285}}.

\bibitem{Gomez:1993ri}
J.~Gomez, R.~Arnold, P.~E. Bosted, C.~Chang, A.~Katramatou, et~al.,
  {Measurement of the A-dependence of deep inelastic electron scattering},
  Phys.Rev. D49 (1994) 4348--4372.
\newblock \href {http://dx.doi.org/10.1103/PhysRevD.49.4348}
  {\path{doi:10.1103/PhysRevD.49.4348}}.

\bibitem{Dasu:1993vk}
S.~Dasu, P.~deBarbaro, A.~Bodek, H.~Harada, M.~Krasny, et~al., {Measurement of
  kinematic and nuclear dependence of R = $\sigma_L / \sigma_T$ in deep
  inelastic electron scattering}, Phys.Rev. D49 (1994) 5641--5670.
\newblock \href {http://dx.doi.org/10.1103/PhysRevD.49.5641}
  {\path{doi:10.1103/PhysRevD.49.5641}}.

\bibitem{Virchaux:1991jc}
M.~Virchaux, A.~Milsztajn, {A Measurement of $\alpha_s$ and higher twists from
  a QCD analysis of high statistics $F_2$ data on hydrogen and deuterium
  targets}, Phys.Lett. B274 (1992) 221--229.
\newblock \href {http://dx.doi.org/10.1016/0370-2693(92)90527-B}
  {\path{doi:10.1016/0370-2693(92)90527-B}}.

\bibitem{Kulagin:2004ie}
S.~A. Kulagin, R.~Petti, {Global study of nuclear structure functions},
  Nucl.Phys. A765 (2006) 126--187.
\newblock \href {http://arxiv.org/abs/hep-ph/0412425}
  {\path{arXiv:hep-ph/0412425}}, \href
  {http://dx.doi.org/10.1016/j.nuclphysa.2005.10.011}
  {\path{doi:10.1016/j.nuclphysa.2005.10.011}}.

\bibitem{Gluck:1993dpa}
M.~Gl{\"u}ck, E.~Reya, M.~Stratmann, {Heavy quarks at high-energy colliders},
  Nucl.Phys. B422 (1994) 37--56.
\newblock \href {http://dx.doi.org/10.1016/0550-3213(94)00131-6}
  {\path{doi:10.1016/0550-3213(94)00131-6}}.

\bibitem{Vermaseren:2005qc}
J.~Vermaseren, A.~Vogt, S.~Moch, {The Third-order QCD corrections to
  deep-inelastic scattering by photon exchange}, Nucl.Phys. B724 (2005) 3--182.
\newblock \href {http://arxiv.org/abs/hep-ph/0504242}
  {\path{arXiv:hep-ph/0504242}}, \href
  {http://dx.doi.org/10.1016/j.nuclphysb.2005.06.020}
  {\path{doi:10.1016/j.nuclphysb.2005.06.020}}.

\bibitem{Ball:2011uy}
R.~D. Ball, et~al., {Unbiased global determination of parton distributions and
  their uncertainties at NNLO and at LO}, Nucl.Phys. B855 (2012) 153--221.
\newblock \href {http://arxiv.org/abs/1107.2652} {\path{arXiv:1107.2652}},
  \href {http://dx.doi.org/10.1016/j.nuclphysb.2011.09.024}
  {\path{doi:10.1016/j.nuclphysb.2011.09.024}}.

\bibitem{Caola:2009iy}
F.~Caola, S.~Forte, J.~Rojo, {Deviations from NLO QCD evolution in inclusive
  HERA data}, Phys.Lett. B686 (2010) 127--135.
\newblock \href {http://arxiv.org/abs/0910.3143} {\path{arXiv:0910.3143}},
  \href {http://dx.doi.org/10.1016/j.physletb.2010.02.043}
  {\path{doi:10.1016/j.physletb.2010.02.043}}.

\bibitem{Lai:2010vv}
H.-L. Lai, M.~Guzzi, J.~Huston, Z.~Li, P.~M. Nadolsky, et~al., {New parton
  distributions for collider physics}, Phys.Rev. D82 (2010) 074024.
\newblock \href {http://arxiv.org/abs/1007.2241} {\path{arXiv:1007.2241}},
  \href {http://dx.doi.org/10.1103/PhysRevD.82.074024}
  {\path{doi:10.1103/PhysRevD.82.074024}}.

\bibitem{Buza:1996wv}
M.~Buza, Y.~Matiounine, J.~Smith, W.~van Neerven, {Charm electroproduction
  viewed in the variable flavor number scheme versus fixed order perturbation
  theory}, Eur.Phys.J. C1 (1998) 301--320.
\newblock \href {http://arxiv.org/abs/hep-ph/9612398}
  {\path{arXiv:hep-ph/9612398}}.

\bibitem{Bierenbaum:2009zt}
I.~Bierenbaum, J.~Bl{\"u}mlein, S.~Klein, {The Gluonic Operator Matrix Elements
  at $O(\alpha_s^2)$ for DIS Heavy Flavor Production}, Phys.Lett. B672 (2009)
  401--406.
\newblock \href {http://arxiv.org/abs/0901.0669} {\path{arXiv:0901.0669}},
  \href {http://dx.doi.org/10.1016/j.physletb.2009.01.057}
  {\path{doi:10.1016/j.physletb.2009.01.057}}.

\bibitem{Bierenbaum:2009mv}
I.~Bierenbaum, J.~Bl{\"u}mlein, S.~Klein, {Mellin Moments of the
  $O(\alpha^3_s)$ Heavy Flavor Contributions to unpolarized Deep-Inelastic
  Scattering at $Q^2 \gg m^2$ and Anomalous Dimensions}, Nucl.Phys. B820 (2009)
  417--482.
\newblock \href {http://arxiv.org/abs/0904.3563} {\path{arXiv:0904.3563}},
  \href {http://dx.doi.org/10.1016/j.nuclphysb.2009.06.005}
  {\path{doi:10.1016/j.nuclphysb.2009.06.005}}.

\bibitem{Ablinger:2010ty}
J.~Ablinger, J.~Bl{\"u}mlein, S.~Klein, C.~Schneider, F.~Wissbrock, {The
  $O(\alpha_s^3)$ Massive Operator Matrix Elements of $O(n_f)$ for the
  Structure Function $F_2(x,Q^2)$ and Transversity}, Nucl.Phys. B844 (2011)
  26--54.
\newblock \href {http://arxiv.org/abs/1008.3347} {\path{arXiv:1008.3347}},
  \href {http://dx.doi.org/10.1016/j.nuclphysb.2010.10.021}
  {\path{doi:10.1016/j.nuclphysb.2010.10.021}}.

\bibitem{Aivazis:1993pi}
M.~Aivazis, J.~C. Collins, F.~I. Olness, W.-K. Tung, {Leptoproduction of heavy
  quarks. 2. A Unified QCD formulation of charged and neutral current processes
  from fixed target to collider energies}, Phys.Rev. D50 (1994) 3102--3118.
\newblock \href {http://arxiv.org/abs/hep-ph/9312319}
  {\path{arXiv:hep-ph/9312319}}, \href
  {http://dx.doi.org/10.1103/PhysRevD.50.3102}
  {\path{doi:10.1103/PhysRevD.50.3102}}.

\bibitem{Thorne:2012qh}
R.~Thorne, {The Effect of Changes of Variable Flavour Number Scheme on PDFs and
  Predicted Cross Sections,~}\href {http://arxiv.org/abs/1201.6180}
  {\path{arXiv:1201.6180}}.

\bibitem{Guzzi:2011ew}
M.~Guzzi, P.~M. Nadolsky, H.-L. Lai, C.-P. Yuan, {General-mass treatment for
  deep inelastic scattering at two-loop accuracy,~}\href
  {http://arxiv.org/abs/1108.5112} {\path{arXiv:1108.5112}}.

\bibitem{Chuvakin:1999nx}
A.~Chuvakin, J.~Smith, W.~van Neerven, {Comparison between variable flavor
  number schemes for charm quark electroproduction}, Phys.Rev. D61 (2000)
  096004.
\newblock \href {http://arxiv.org/abs/hep-ph/9910250}
  {\path{arXiv:hep-ph/9910250}}, \href
  {http://dx.doi.org/10.1103/PhysRevD.61.096004}
  {\path{doi:10.1103/PhysRevD.61.096004}}.

\bibitem{Lipka}
{{Lipka,K.,~}}, {{these proceedings.}}

\bibitem{Forte:2010ta}
S.~Forte, E.~Laenen, P.~Nason, J.~Rojo, {Heavy quarks in deep-inelastic
  scattering}, Nucl.Phys. B834 (2010) 116--162.
\newblock \href {http://arxiv.org/abs/1001.2312} {\path{arXiv:1001.2312}},
  \href {http://dx.doi.org/10.1016/j.nuclphysb.2010.03.014}
  {\path{doi:10.1016/j.nuclphysb.2010.03.014}}.

\bibitem{Aad:2011yn}
G.~Aad, et~al., {Measurement of the Muon Charge Asymmetry from W Bosons
  Produced in pp Collisions at $\sqrt{s} = 7$ TeV with the ATLAS detector},
  Phys.Lett. B701 (2011) 31--49.
\newblock \href {http://arxiv.org/abs/1103.2929} {\path{arXiv:1103.2929}},
  \href {http://dx.doi.org/10.1016/j.physletb.2011.05.024}
  {\path{doi:10.1016/j.physletb.2011.05.024}}.

\bibitem{Chatrchyan:2011jz}
S.~Chatrchyan, et~al., {Measurement of the lepton charge asymmetry in inclusive
  $W$ production in pp collisions at $\sqrt{s} = 7$ TeV}, JHEP 1104 (2011) 050.
\newblock \href {http://arxiv.org/abs/1103.3470} {\path{arXiv:1103.3470}},
  \href {http://dx.doi.org/10.1007/JHEP04(2011)050}
  {\path{doi:10.1007/JHEP04(2011)050}}.

\bibitem{Catani:2010en}
S.~Catani, G.~Ferrera, M.~Grazzini, {W boson production at hadron colliders:
  the lepton charge asymmetry in NNLO QCD}, JHEP 1005 (2010) 006.
\newblock \href {http://arxiv.org/abs/1002.3115} {\path{arXiv:1002.3115}},
  \href {http://dx.doi.org/10.1007/JHEP05(2010)006}
  {\path{doi:10.1007/JHEP05(2010)006}}.

\bibitem{Alekhin:2000es}
S.~I. Alekhin, {Statistical properties of the estimator using covariance
  matrix,~}\href {http://arxiv.org/abs/hep-ex/0005042}
  {\path{arXiv:hep-ex/0005042}}.

\bibitem{Alekhin:2011ey}
S.~Alekhin, J.~Bl{\"u}mlein, S.~Moch, {Higher order constraints on the Higgs
  production rate from fixed-target DIS data}, Eur.Phys.J. C71 (2011) 1723.
\newblock \href {http://arxiv.org/abs/1101.5261} {\path{arXiv:1101.5261}},
  \href {http://dx.doi.org/10.1140/epjc/s10052-011-1723-1}
  {\path{doi:10.1140/epjc/s10052-011-1723-1}}.

\bibitem{Ball:2011us}
R.~D. Ball, V.~Bertone, L.~Del~Debbio, S.~Forte, A.~Guffanti, et~al.,
  {Precision NNLO determination of $\alpha_s(M_Z)$ using an unbiased global
  parton set}, Phys.Lett. B707 (2012) 66--71, long author list - awaiting
  processing.
\newblock \href {http://arxiv.org/abs/1110.2483} {\path{arXiv:1110.2483}},
  \href {http://dx.doi.org/10.1016/j.physletb.2011.11.053}
  {\path{doi:10.1016/j.physletb.2011.11.053}}.

\bibitem{Bolzoni:2010xr}
P.~Bolzoni, F.~Maltoni, S.~Moch, M.~Zaro, {Higgs production via vector-boson
  fusion at NNLO in QCD}, Phys.Rev.Lett. 105 (2010) 011801.
\newblock \href {http://arxiv.org/abs/1003.4451} {\path{arXiv:1003.4451}},
  \href {http://dx.doi.org/10.1103/PhysRevLett.105.011801}
  {\path{doi:10.1103/PhysRevLett.105.011801}}.

\bibitem{GehrmannDeRidder:2011aa}
A.~Gehrmann-De~Ridder, E.~Glover, J.~Pires, {Real-Virtual corrections for gluon
  scattering at NNLO,~}\href {http://arxiv.org/abs/1112.3613}
  {\path{arXiv:1112.3613}}.

\bibitem{Bolzoni:2010bt}
P.~Bolzoni, G.~Somogyi, Z.~Trocsanyi, {A subtraction scheme for computing QCD
  jet cross sections at NNLO: integrating the iterated singly-unresolved
  subtraction terms}, JHEP 1101 (2011) 059.
\newblock \href {http://arxiv.org/abs/1011.1909} {\path{arXiv:1011.1909}},
  \href {http://dx.doi.org/10.1007/JHEP01(2011)059}
  {\path{doi:10.1007/JHEP01(2011)059}}.

\bibitem{Alekhin:2011cf}
S.~Alekhin, J.~Bl{\"u}mlein, S.~Moch, {Parton distributions and Tevatron jet
  data,~}\href {http://arxiv.org/abs/1105.5349} {\path{arXiv:1105.5349}}.

\bibitem{Nagy:2001fj}
Z.~Nagy, {Three jet cross-sections in hadron hadron collisions at
  next-to-leading order}, Phys.Rev.Lett. 88 (2002) 122003.
\newblock \href {http://arxiv.org/abs/hep-ph/0110315}
  {\path{arXiv:hep-ph/0110315}}, \href
  {http://dx.doi.org/10.1103/PhysRevLett.88.122003}
  {\path{doi:10.1103/PhysRevLett.88.122003}}.

\bibitem{Nagy:2003tz}
Z.~Nagy, {Next-to-leading order calculation of three jet observables in hadron
  hadron collision}, Phys.Rev. D68 (2003) 094002.
\newblock \href {http://arxiv.org/abs/hep-ph/0307268}
  {\path{arXiv:hep-ph/0307268}}, \href
  {http://dx.doi.org/10.1103/PhysRevD.68.094002}
  {\path{doi:10.1103/PhysRevD.68.094002}}.

\bibitem{Kidonakis:2000gi}
N.~Kidonakis, J.~Owens, {Effects of higher order threshold corrections in high
  E(T) jet production}, Phys.Rev. D63 (2001) 054019.
\newblock \href {http://arxiv.org/abs/hep-ph/0007268}
  {\path{arXiv:hep-ph/0007268}}, \href
  {http://dx.doi.org/10.1103/PhysRevD.63.054019}
  {\path{doi:10.1103/PhysRevD.63.054019}}.

\bibitem{Kluge:2006xs}
T.~Kluge, K.~Rabbertz, M.~Wobisch, {FastNLO: Fast pQCD calculations for PDF
  fits,~}\href {http://arxiv.org/abs/hep-ph/0609285}
  {\path{arXiv:hep-ph/0609285}}.

\bibitem{Wobisch:2011ij}
M.~Wobisch, D.~Britzger, T.~Kluge, K.~Rabbertz, F.~Stober, {Theory-Data
  Comparisons for Jet Measurements in Hadron-Induced Processes,~}\href
  {http://arxiv.org/abs/1109.1310} {\path{arXiv:1109.1310}}.

\bibitem{Abulencia:2007ez}
A.~Abulencia, et~al., {Measurement of the Inclusive Jet Cross Section using the
  {\boldmath $k_{\rm T}$} algorithmin{\boldmath $p\overline{p}$} Collisions
  at{\boldmath $\sqrt{s}$} = 1.96 TeV with the CDF II Detector}, Phys.Rev. D75
  (2007) 092006.
\newblock \href {http://arxiv.org/abs/hep-ex/0701051}
  {\path{arXiv:hep-ex/0701051}}, \href
  {http://dx.doi.org/10.1103/PhysRevD.75.092006, 10.1103/PhysRevD.75.119901}
  {\path{doi:10.1103/PhysRevD.75.092006, 10.1103/PhysRevD.75.119901}}.

\bibitem{Aaltonen:2008eq}
T.~Aaltonen, et~al., {Measurement of the Inclusive Jet Cross Section at the
  Fermilab Tevatron p anti-p Collider Using a Cone-Based Jet Algorithm},
  Phys.Rev. D78 (2008) 052006.
\newblock \href {http://arxiv.org/abs/0807.2204} {\path{arXiv:0807.2204}},
  \href {http://dx.doi.org/10.1103/PhysRevD.78.052006,
  10.1103/PhysRevD.79.119902} {\path{doi:10.1103/PhysRevD.78.052006,
  10.1103/PhysRevD.79.119902}}.

\bibitem{Abazov:2008hua}
V.~Abazov, et~al., {Measurement of the inclusive jet cross-section in $p
  \bar{p}$ collisions at $s^{91/2)}$ =1.96-TeV}, Phys.Rev.Lett. 101 (2008)
  062001.
\newblock \href {http://arxiv.org/abs/0802.2400} {\path{arXiv:0802.2400}},
  \href {http://dx.doi.org/10.1103/PhysRevLett.101.062001}
  {\path{doi:10.1103/PhysRevLett.101.062001}}.

\bibitem{Abazov:2010fr}
V.~Abazov, et~al., {Measurement of the dijet invariant mass cross section in
  proton anti-proton collisions at sqrt{s} = 1.96 TeV}, Phys.Lett. B693 (2010)
  531--538.
\newblock \href {http://arxiv.org/abs/1002.4594} {\path{arXiv:1002.4594}},
  \href {http://dx.doi.org/10.1016/j.physletb.2010.09.013}
  {\path{doi:10.1016/j.physletb.2010.09.013}}.

\bibitem{Alekhin:2010dd}
S.~Alekhin, J.~Bl{\"u}mlein, P.~Jimenez-Delgado, S.~Moch, E.~Reya, {NNLO
  Benchmarks for Gauge and Higgs Boson Production at TeV Hadron Colliders},
  Phys. Lett. B697 (2011) 127--135.
\newblock \href {http://arxiv.org/abs/1011.6259} {\path{arXiv:1011.6259}},
  \href {http://dx.doi.org/10.1016/j.physletb.2011.01.034}
  {\path{doi:10.1016/j.physletb.2011.01.034}}.

\bibitem{Rabbertz:1368241}
K.~Rabbertz, Comparison of inclusive jet and dijet mass cross sections at
  $\sqrt{s}$ = 7tev with predictions of perturbative qcd, Tech. Rep.
  CMS-NOTE-2011-004., CERN, Geneva (Jun 2011).

\bibitem{Aad:2011fc}
G.~Aad, et~al., {Measurement of inclusive jet and dijet production in pp
  collisions at sqrt(s) = 7 TeV using the ATLAS detector,~}\href
  {http://arxiv.org/abs/1112.6297} {\path{arXiv:1112.6297}}.

\bibitem{Wobisch:2012iu}
M.~Wobisch, {Recent QCD results from the Tevatron,~}\href
  {http://arxiv.org/abs/1202.0205} {\path{arXiv:1202.0205}}.

\bibitem{Amhis:2012gj}
Y.~Amhis, {Electroweak results at LHCb,~}Presented at the 2011 Hadron Collider
  Physics symposium (HCP-2011), Paris, France, November 14-18 2011, 4 pages, 8
  figures.
\newblock \href {http://arxiv.org/abs/1202.0654} {\path{arXiv:1202.0654}}.

\end{thebibliography}

\end{document}